\documentclass[useAMS,usenatbib]{mn2e}
\usepackage{graphicx}
\usepackage{enumerate}
\usepackage[version=3]{mhchem} 
\usepackage{subfig}
\usepackage{lscape}
\usepackage{epstopdf}
\usepackage{amssymb}
\usepackage{color,ulem}

\title{Size and density sorting of dust grains in SPH simulations of protoplanetary discs}
\author[Pignatale et al.]{F. C. Pignatale$^{1}$\thanks{E-mail:
francesco.pignatale@univ-lyon1.fr}, J.-F. Gonzalez$^{1}$, Nicolas Cuello$^{1,2,3}$, Bernard Bourdon$^{4}$, Caroline Fitoussi$^{4}$ \\
$^{1}$ 
Univ Lyon, Univ Lyon1, Ens de Lyon, CNRS, Centre de Recherche Astrophysique de Lyon UMR5574, F-69230, Saint-Genis-Laval, France\\
$^{2}$ Instituto de Astrof\'{i}sica, Pontificia Universidad Cat\'{o}lica de Chile, Santiago, Chile\\
$^{3}$ Millennium Nucleus ``Protoplanetary Disks'', Chile, Chile\\
$^{4}$ 	Univ Lyon, Ens de Lyon, Universit\'e Lyon 1, CNRS, UMR 5276 LGL-TPE, F-69342, Lyon, France
  \\
}

\begin{document}

\date{Accepted 2017 March 30. Received 2017 March 30; in original form 2016 May 10.}

\pagerange{\pageref{firstpage}--\pageref{lastpage}} \pubyear{2015}

\maketitle

\label{firstpage}

\begin{abstract}

The size and density of dust grains determine their response to  gas drag in protoplanetary discs. Aerodynamical (size$\times$density) sorting is one of the proposed mechanisms to explain the grain properties and chemical fractionation of chondrites. However, the efficiency of aerodynamical sorting and the location in the disc in which it could occur are still unknown. Although the effects of grain sizes and growth in discs have been widely studied, a simultaneous analysis including dust composition is missing. In this work we present the dynamical evolution and growth of multicomponent dust in a protoplanetary disc using a 3D, two-fluid (gas+dust)  Smoothed Particle Hydrodynamics (SPH) code.

We find that the dust vertical settling is characterised by two phases: a  density-driven phase which leads to a vertical  chemical sorting of dust and a size-driven phase which  enhances the amount of lighter material in the midplane. We also see an efficient radial chemical sorting of the dust at large scales. We find that dust particles are aerodynamically sorted in the inner disc. The disc becomes sub-solar in its \ce{Fe}/\ce{Si} ratio on the surface since the early stage of evolution but sub-solar \ce{Fe}/\ce{Si} can be also found in the outer disc-midplane at late stages.

Aggregates in the disc mimic the physical and chemical properties of chondrites, suggesting that aerodynamical sorting played an important role in determining their final structure.

\end{abstract}


\section{Introduction}
\label{intro}

In protoplanetary discs the interaction between gas and dust plays a central role in determining the dust dynamics. The settling and drift of a dust particle due to the aerodynamic drag exerted by the surrounding gas is driven by the size and the density of the considered particle and, thus, by the stopping time,
\begin{equation}
t_{\rm s} = \frac{\rho_{\rm d} s_{\rm d} }{c_{\rm s} \rho_{\rm g}} ,\
\label{stoppingtime}
\end{equation}
where $s_{\rm d}$ is the size (radius) of the particle, $\rho_{\rm d}$ its intrinsic density, $c_{\rm s}$ the sound speed, and $\rho_{\rm g}$ the gas density \citep{1977MNRAS.180...57W}. As such, for a given set of parameters ($c_{\rm s}, \rho_{\rm g}$), the dynamical behaviour of the particle is determined by the product of size and density, which is called the aerodynamic parameter \citep{2006mess.book..353C}, $\zeta=\rho_{\rm d} s_{\rm d}$ : smaller and lighter particles have short stopping times and are well coupled to the gas while larger and denser grains have longer stopping times and are less coupled to the surrounding gas. 

Furthermore, the  rate of radial drift of a particle reaches its maximum when 
$t_{\rm s}/t_{\rm orb}=1/2\pi$  \citep{1977MNRAS.180...57W}, where $t_{\rm orb}$ is the orbital period
\begin{equation}
t_{\rm orb} =  \frac{2\pi}{\Omega_k} \,.
\label{tormb}
\end{equation}
Thus, combining the stopping time, equation~(\ref{stoppingtime}), with the orbital period, the optimal drift size for a given species, corresponding to the fastest drift rate, $s_{\rm opt}$, can be derived:
\begin{equation}
s_{\rm opt} =  \frac{c_{\rm s} \rho_{\rm g}}{\Omega_k \rho_{\rm d}} \,,
\label{optimalone}
\end{equation}
\citep{2007A&A...474.1037F,2008A&A...487..265L}. Equation~\ref{optimalone} states that for given  conditions, the optimal drift size varies with the  intrinsic density of the considered dust particle. Moreover, the value of $s_{\rm opt}$ in the midplane, $s_{\rm opt}^{\rm mid}$,  can be written as 
\begin{equation}
s_{\rm opt}^{\rm mid} =  \frac{\Sigma_{\rm g}}{\sqrt{2\pi} \rho_{\rm d}} \,,
\label{optimalone2}
\end{equation}
\citep{2010A&A...518A..16F}, where $\Sigma_{\rm g}$ is the gas surface density. The dependence of $s_{\rm opt}^{\rm mid}$ on the particle intrinsic density is also evident.

The different response to the gas drag of dusty aggregates with different sizes and densities is thought to have contributed to the determination of the physical and chemical properties of the pristine complex aggregates in our early Solar System: chondrites \citep{2003TrGeo...1..143S,2006mess.book..353C}. Chondrites are undifferentiated meteorites, which are characterized by the presence of chondrules, small spherules 0.01-10~mm in size \citep{2003TrGeo...1..143S}. Chondrites host other components: (i) calcium-aluminum rich inclusion  (CAIs), made of refractory material, (ii) amoeboid olivine aggregates, characterized by fine olivine grains, metals and refractory compounds, (iii) metallic grains (Fe-Ni), and (iv) matrix, an unequilibrated volatile-rich mixture \citep{2003TrGeo...1..201M,2003TrGeo...1..143S,2007AREPS..35..577S}. Chondrites are among the oldest rocks known. They retain radiometric ages corresponding to the first few Myr after calcium--aluminium rich inclusions (whose formation dates at 4567.2 $\pm$0.6 Myr ago \citep{2002Sci...297.1678A}), and thus record events that occurred at the beginning of the Solar System formation. 

\begin{figure}
{\includegraphics[width=1\columnwidth]{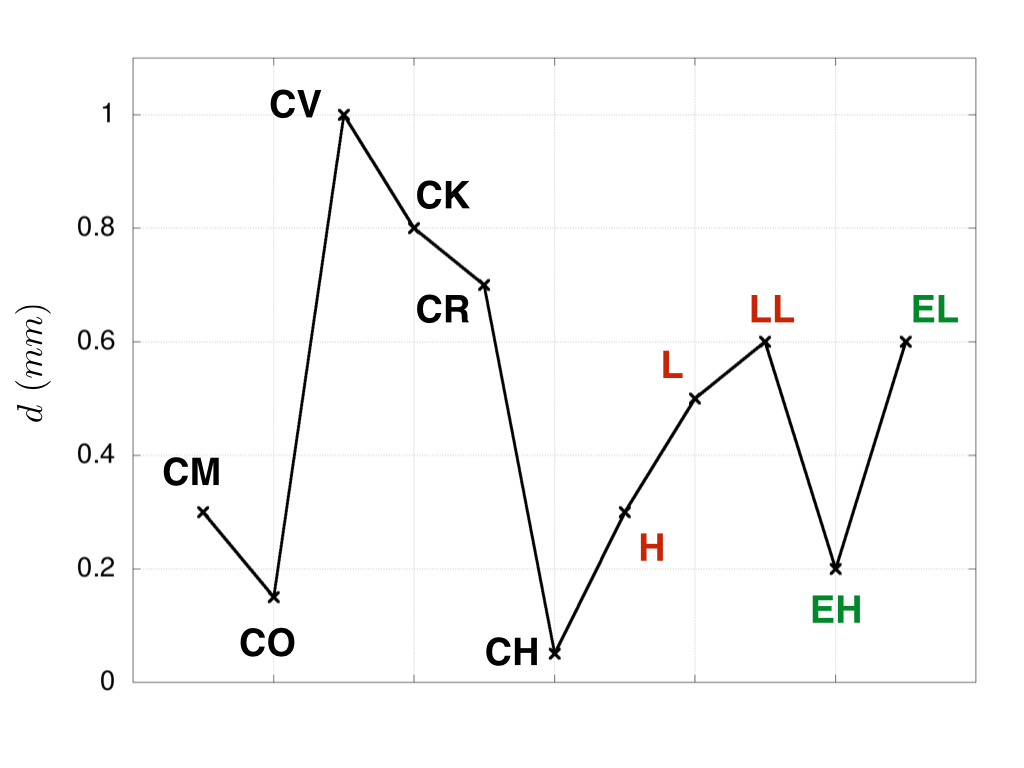}} \\
\caption{Average diameter of chondrules in different groups of chondrites. Figure adapted  from  \citet{Palme200341} and \citet{2005ASPC..341...15S}. \label{chondsize}}
\end{figure}

The formation of chondrules was an ongoing process which occurred from the formation of CAIs and lasted ~3 Myr \citep{2012Sci...338..651C}. The size of chondrules varies within each clan as well as between  different chondrite groups \citep{2010GeCoA..74.4807R}, see Fig.~\ref{chondsize}. 
There is a general size-density correspondence between chondrules, metallic grains (Fe-Ni) and sulfide grains, within the same group of chondrites, for which $(\rho s)_{\rm chond}\sim(\rho s)_{\rm metal}\sim(\rho s)_{\rm sulf}$ \citep{1998LPI....29.1457B,1999Icar..141...96K}. Size sorting and size-density sorting due to gas drag in different scenarios (local turbulences, large disc scale turbulences, jet-flow) are some of the  proposed mechanisms to explain the chondrule size distribution, the assemblage of the chondrite components,  and the size-density correspondence between chondrules, metallic grains (Fe-Ni) and sulfide grains, within the same group of chondrites  \citep{1980E&PSL..47..199C,1998LPI....29.1457B,
1999Icar..141...96K,2001ApJ...546..496C,2005M&PS...40..123L,2006E&PSL.248..650Z,
2012Icar..220..162J}.

\begin{figure}
{\includegraphics[width=1\columnwidth]{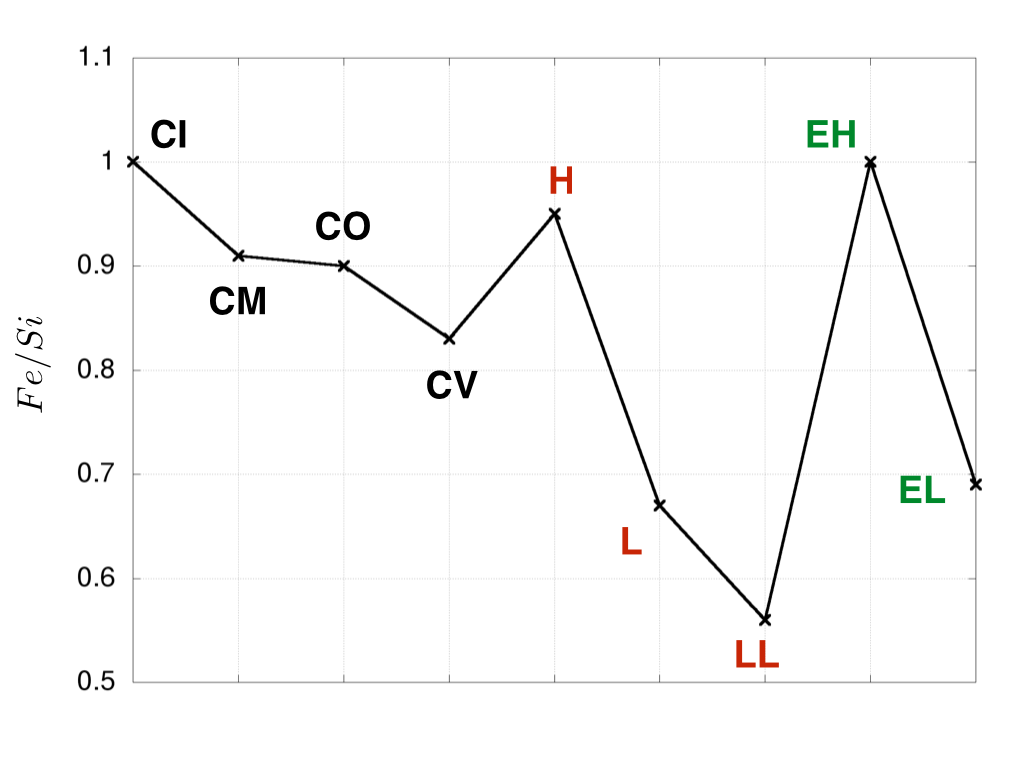}} \\
\caption{\ce{Fe}/\ce{Si} ratios for different groups of chondrites normalized to CI chondrites, which are assumed to be the solar standard. This figure is adapted from \citet{Palme200341}. \label{fractionation}}
\end{figure}

Chondrites exhibit different degrees  of elemental fractionation, for example in their metal-silicate content \citep{1970GeCoA..34..367L}. The \ce{Fe}/\ce{Si} ratio  of chondrites normalized to CI chondrites (whose \ce{Fe}/\ce{Si} ratio is close to the solar value) varies according to the chondrites group \citep{Palme200341,2005ASPC..341...15S}, see Fig.~\ref{fractionation}. 
\begin{figure}
{\includegraphics[width=1\columnwidth]{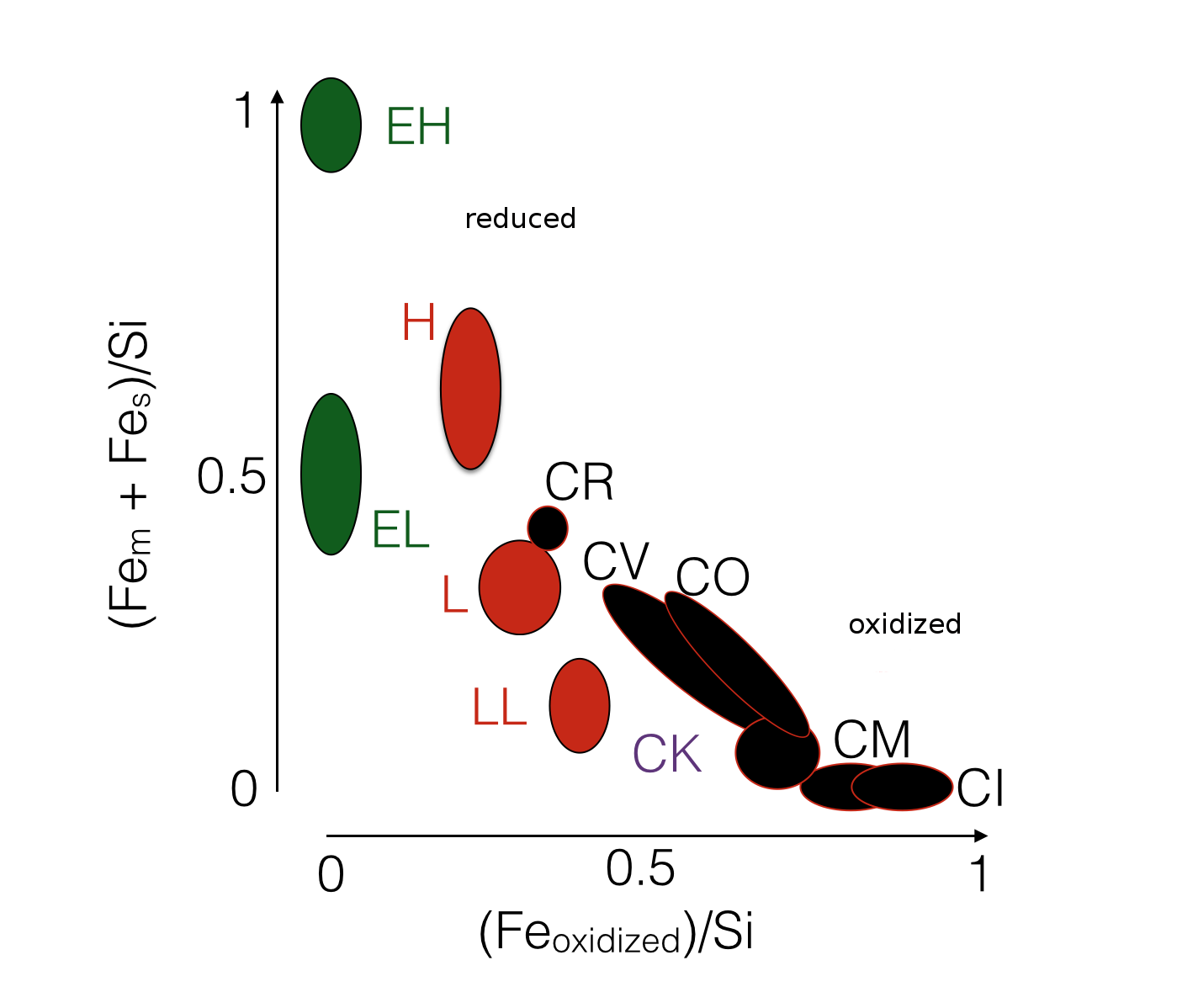}} \\
\caption{Urey-Craig  diagram illustrating the ratio of oxidized versus metal+sulfide iron to silicon, in different groups of chondrites. Values are expressed in mole ratio. This diagram is adapted from \citet{2006mess.book..803R}. \label{urey}}
\end{figure}
Moreover the ratio of oxidized iron and metallic iron+sulfides to silicon, changes for different chondrite groups, Fig.~\ref{urey}. The chemical and physical properties of chondrites suggest that chondrites formed, accreted and evolved in distinct disc regions and at different times, within different chemical reservoirs \citep{2005ASPC..341...15S}.

Aerodynamic sorting, photophoresis, magnetic properties, grain aggregation, and heating processes, both in large (disc, eddies) and small (chondritic parent bodies) scales, are among the suggested theories to produce the observed trend \citep{1988mess.book..394L, 1988mess.book..416L,1999Icar..141...96K, 2003GeoRL..30.1522M,2013ApJ...769...78W,2016MNRAS.458.2140C}. However, the level of  contribution of dynamical sorting  in assembling the chondritic material is not well constrained \citep{2010GeCoA..74.4807R}.

Size sorting of dust has been observed in protoplanetary discs. \citet{2007A&A...469..963P} modeled the dust size distribution for GG~Tau, a circumbinary disc, finding that multiple populations of dust, each with different sizes and scale heights, are necessary to match the observed brightness profile. They found that 1~$\mu$m, 2.5$\mu$m, 5$\mu$m, 7.5$\mu$m, 10$\mu$m grains have a decreasing scale height suggesting a stratified structure of the dust. Another T-Tauri object which shows dust sorting is TW~Hya. TW~Hya has been widely studied at different wavelengths, each probing different dust sizes \citep{2000ApJ...534L.101W,2003ApJ...596..597W,2007ApJ...664..536H,
2008ApJ...678.1119H,2012ApJ...744..162A,2014A&A...564A..93M}. All observations suggest that  vertical and radial sorting of the dust is efficient in  TW~Hya: the abundance of larger grains peaks at the midplane and in the inner radial regions {\rm while smaller grains encompass the larger grains occupying large radial distances and  vertical extension of the disc}. Thus, observation suggests that sorting is a process which characterises the dust dynamics at very large disc scales.

Several studies in the past investigated the vertical settling and radial drift of grains as a function of their size \citep{2004A&A...421.1075D,2005A&A...443..185B,2008A&A...487..205D,2009Icar..200..655C},   and in the resulting observational properties of discs \citep{2004A&A...421.1075D}. An extensive review of the processes involved in  gas-dust dynamics and interactions has been made by \citet{2006mess.book..353C}.

In addition, the early stage of planet formation is characterized by a fast growth of submicron-size grains to large planetesimals \citep{2007prpl.conf..783D}.  Signatures of rapid grain growth have been extensively found in protoplanetary discs from observations at different wavelenghts \citep{2001Sci...292.1686T,2003A&A...403..323T,2008ApJ...683..479B,2012MNRAS.425.3137U}.
The growth of grains is a complex process which involves, beside the physical conditions of the disc, the chemical  and physical properties of the dust \citep{2007A&A...461..215O,2008ARA&A..46...21B,2012ApJ...752..106O}. Moreover, collisions of grains do not always produce perfect sticking and fragmentation and bouncing of grains upon collisions can occur \citep{2005A&A...434..971D,2010A&A...513A..57Z}. Grain growth in protoplanetary discs has been the subject of theoretical \citep{1997A&A...319.1007S,2001ApJ...551..461S,2005A&A...434..971D,2005MNRAS.362.1015H,
2007ApJ...654L.159C,2008A&A...487..265L,2010A&A...520A..43O,2012A&A...539A.148B}, and experimental investigation \citep{2000PhRvL..85.2426B,2000Icar..143..138B,2008ARA&A..46...21B,2009MNRAS.393.1584T}.

However, to date, there is no complete full 3D analysis of the dust dynamics (vertical settling and radial drift)  of a chemically heterogeneous dust mixture. The aim of this work is, thus, to simultaneously investigate the effects of grain growth, size and density  in shaping the dust distribution of a multicomponent dust mixture at large disc scales, and  to gain more insight into the dynamical processes which may have contributed to assembling chondritic material and the building blocks of planets. 

\section{Methods}
\label{method}
We compute the vertical settling, radial drift and growth of a multicomponent dust using our 3D, two-fluid  (gas+dust) SPH code \citep{2005A&A...443..185B,2008A&A...487..265L}. In this section we describe the structure of our chosen protoplanetary disc, section~\ref{Disc}, the adopted grain chemistry, section~\ref{chemistry}, the prescription used for grain growth, section~\ref{growthpresc}, and detail our simulations, section~\ref{simulations}. 

\subsection{Disc Model}
\label{Disc}

 In order to make comparisons with previous work, we consider  a typical T-Tauri disc used in \citet{2005A&A...443..185B} and \citet{2008A&A...487..265L} as our fiducial disc model, which represents the initial state of our simulations. The mass of the central star is $M_{\star}=1~M_{\odot}$,  and the disc is described by the following parmeters: the mass of the disc is $M_{\rm disc}=0.02~M_{\star}$ with a radial extension of $20\le R\rm{(au)} \le 400$. The disc is composed of 99$\%$  gas and 1$\%$  dust by mass. The parametrization for the temperature reads as $T \propto R^{-3/4}$, and for the surface density as  $\Sigma \propto R^{-3/2}$. The disc is vertically isothermal and the sound speed is $c_{\rm s} \propto R^{-3/8}$. The vertical scale height is $H = c_{\rm s}/\Omega_{\rm K}$, and $H/R \propto R^{1/8}$. The disc is flared and $H/R =0.05$ at our reference radius, $R_0 = 100$~AU. For all simulations, the artificial viscosity is set to $\alpha_{\rm SPH}=0.1$ and $\beta_{\rm SPH}=0.0$ emulating a viscosity parameter $\alpha_{\rm ss}=0.01$ \citep{1973A&A....24..337S,2007A&A...474.1037F}, to reproduce the observed stellar accretion rate \citep{2007MNRAS.376.1740K}.

\subsection{Grain Chemistry}
\label{chemistry}
 		
\begin{table}							
\begin{tabular} {l| c| c|c|}							
Group	&	dust species	&	wt$\%$	&	 density 	\\
	&		&		&	$\rm{g cm^{-3}}$	\\
\hline							
\hline							
Fe	&	metallic iron	&	1	&	7.87	\\
	&	troilite	&	6	&	4.83	\\
\hline							
Si	&	olivine \& 	&	25	&	3.46	\\
	&	pyroxene	&		&		\\
\hline							
org	&	refractory organics	&	25	&	1.5	\\
	&	volatile organics	&	4	&	1	\\
\hline							
\ce{H2O}	&	ice	&	39	&	0.92	\\
\hline							
\hline	
		&	 	&  ratios	&		\\
\hline											
\ce{H2O}/Si	&		&	1.56	&		\\
Fe/Si	&		&	0.28	&		\\
\hline																			
\end{tabular}							
\caption{Dust distribution from \citet{1994ApJ...421..615P} describing the chemical composition for large discs with  related wt$\%$ and intrinsic densities. Ratios between groups of different species are also shown. \label{frac-abundance}}	
\end{table}

Dust in discs is a complex mixture of several species. The chemistry of the dust is strictly related to the physical conditions in each zone of the disc and to the composition of the gas and the pristine dust \citep{1998A&A...332.1099G}. In the hotter inner zone of the disc, dust experiences several chemical processes such as direct condensation from the gas phase, annealing, melting, shocks, and gas-grain surface interaction \citep{2013ChRv..113.9016H}. In the outer cooler part of the disc, the low temperatures and slow kinetics prevent all these processes to occur efficiently and the majority of the dust will generally keep its pristine chemical composition and structure for longer time-scales. Our disc extends from $20\le R\rm{(AU)} \le 400$, thus we do not take in account any chemical transformation of the dust grains. Furthermore, since the inner limit of our disc is beyond  the major ice lines (\ce{H_2O}, \ce{CO}), we do not consider any change of state  (i.e. vaporisation of ice particles or condensation from the gas phase). Indeed the location of the \ce{H_2O}-ice line is well within our inner disc edge ($R_{\rm in}$=20~au) when using the condensation temperatures of 145 K and  170 K for \ce{H_2O}-ice reported in \citet{2012MNRAS.425L...6M}. The approximate location of the \ce{CO}-ice line is $16\le R{\rm(au)}\le21,5$  using the condensation temperatures  of 25 K and 20 K for \ce{CO} reported in \citet{1993prpl.conf.1177M} and  \citet{1993Icar..106..478S}. As such we can neglect the effect\footnote{The occurrence of ice lines could change the chemical composition of the dust, by removing or adding components, and alter the dust size distribution of grains. Nevertheless, the overall dynamical behaviour of grains will still be ruled by their resulting aerodynamic parameter, $\zeta$.  } of these ice-lines on the dust dynamics.

The dust mixture in our disc is taken from \citet{1994ApJ...421..615P} which is representative of the dust mixture present in the outer region of protoplanetary discs. The dust is composed of water ice (\ce{H2O}), volatile organics (\ce{CH3OH}, \ce{H2CO}, $(\ce{H2CO})_x$), refractory organics (\ce{CHON}), a mixture of silicates (olivines and  pyroxenes, with Fe/(Fe+Mg)=0.3), sulfide-rich grains (namely \ce{FeS}) and iron-rich grains (namely \ce{Fe}), with their  values of intrinsic density, $\rho_{\rm d}$, reported in Table~\ref{frac-abundance} together with their weight percent (wt$\%$).

In Table~\ref{frac-abundance} we also report  some dust ratios which will be used during our discussion.  They represent the  ratios between   the wt$\%$ of two given groups in a well mixed disc and not the chemical ratios. In this paper we refer to these values as ``solar" for ease of reading.

In order to account for the different chemical species, we modified the SPH code described in \citet{2005A&A...443..185B} by introducing the possibility to assign different intrinsic density values to different particles of the dust fluid. The user can now choose a number of different dust species and the amount (wt$\%$) of each species present in the disc. 
When the dust is injected in the system the SPH dust-particles are assigned with an intrinsic density according the percentage distribution. In the present work, we follow the distribution reported in Table~\ref{frac-abundance}.  A SPH dust particle with a given $\rho_{\rm d}$ would represent a packet of physical dust particles whose intrinsic density strongly peaks to  $\rho_{\rm d}$ and where the amount of other species is irrelevant. This approach is similar to that introduced in \citet{2008A&A...487..265L} for describing the  size  distribution of growing particles.

\subsection{Grain Growth}
\label{growthpresc}

Observations of pristine dust in protoplanetary discs are characterized by aggregates  with sizes of the order of sub-microns to microns \citep{2014prpl.conf..339T}. In our simulations the initial grain size, $s_{0}$, is common for all particles, and is set to $10\mu \rm m$. \citet{2008A&A...487..265L}  show that the results of their simulations with grain growth have little dependence on $s_{0}$ when the initial size is in the sub-micron to micron range.

Aggregates in discs are made of different pristine components, which we call grains. In section~\ref{chemistry} we assumed that our initial grains comprise the species in Table~1 with their relative abundances \citep{1994ApJ...421..615P}. Grain growth is a complex process, and as first order approximation one could assume that growth occurs as a result of random collision between components of the same or different species. The resulting aggregates would be characterized by a distribution of densities which ranges between the two extremes we consider here (iron and ice), with populations of iron-, silicate- and ice-enriched grains.  Due to the initial abundance differences of our species, the resulting aggregates will  maintain a spread in their densities,  as it is unlikely to produce only one single final group of aggregates characterised by the same type of mixture.

Dynamics dictates how aggregates will evolve in a disc. \citet{2008A&A...487..265L} and  \citet{2014MNRAS.437.3037L,2014MNRAS.437.3025L,2014MNRAS.437.3055L} investigated in detail the differences in the timescales between vertical settling and growth rates, and between radial drift and growth rates. They  concluded that the vertical settling is almost one hundred times faster than the radial drift, and that these two dynamical events can be considered as separate. Moreover, grains grow moderately during the settling phase compared to their final size  ($s=100~\mu \rm m$, at 100~(au), see appendix~\ref{densvsgrowth}) and the efficiency of growth increases once the grains settle in the midplane.

Since the initial size of our grains, $s_{0}$, is equal and common for all particles ($10~\mu \rm m$), grains with different compositions will behave differently as their stopping time is a function of their density (see equation~\ref{stoppingtime}). Given the moderate growth of grains during vertical settling, the settling can result in inhomogeneities  in the chemical content of the disc.  In the midplane,  the radial drift is then dictated by $\zeta$ \citep{1977MNRAS.180...57W}. Aggregates of the same size but different compositions will drift at different velocities. These considerations suggest that grain dynamics in the disc should, in fact, enhance chemical inhomogeneity. Evidences from meteorites  also point to the presence in the Solar Nebula of different chemical reservoirs  from which they accreted \citep{Palme200341,2005ASPC..341...15S}. The verification of these predictions and their evolution with time are some of the goals of this work.

Grain growth  in our SPH code follows  \citet{2008A&A...487..265L}. Grains within an SPH dust particle are assumed to stick perfectly upon collision. Our prescription for  grain growth is taken from \citet{1997A&A...319.1007S}, where $\mathrm{d}s/\mathrm{d}t$ is described by the following equation
\begin{equation}
\label{stepval}
\frac{{\rm d}s}{{\rm d}t}= \sqrt{2^{2/3}{\rm R_{0}}\alpha} \frac{\hat{\rho}}{\rho_{\rm d}} c_{\rm s} \frac{\sqrt{{\rm Sc}-1}}{\rm Sc} ,\
\end{equation}
where  $R_{0}$ is the Rossby number, $\alpha$ the \citet{1973A&A....24..337S} turbulence parameter, $\hat{\rho}$  the density of matter concentrated into solid particles, $\rho_{\rm d}$ the intrinsic density of the physical grains and ${\rm Sc}$ the Schmidt number.

In our simulations when aggregates grow they keep their intrinsic density constant, i.e. there is no mixed growth between particles of different species. This as a direct consequence of the growth prescription  of \citet{1997A&A...319.1007S} which assumes collisions occur each time between two identical (size and density) spherical particles with growth occuring within a SPH particle \citep{2008A&A...487..265L}.

While it may appear non-physical to only consider growth between the same species, this simplified assumption does mimic the true density distribution of real growth between species of mixed composition. The initial difference in the abundances of components-grains and their dynamical dependence on density suggest that we will have, at any given time and location, aggregates with different compositions. Aggregates whose resulting density is iron-enriched will behave similarly to our ``iron grains'', while aggregates which are characterized by a high ice content will behave similarly to our ``ice grains'', and silicate-rich  grains will behave similarly to our ``pure silicates''. As such, having a fixed discrete distribution of $\rho_{\rm d}$ between two extremes (ice and iron) and four intermediate species  will allow us to trace and investigate the different behaviours driven by the intrinsic density and the aerodynamic parameter. Any interpolation of the behaviour of aggregates with intermediate densities is then straightforward. In fact, this approach is similar to the widely used approximation to study the effect of  size on dust dynamics for which  a discrete grain size distribution and often one chemical species is considered \citep{2005A&A...443..185B,2008A&A...487..205D,2010ApJ...722.1437B,2012ApJ...753..119C}. These simulations proved to be in very good agreement with observations. Our simulations, instead, account  for a discrete distribution of densities and a time-evolving distribution of size since we also include grain growth. Growth can erase the illustrated effect of densities only if, at any given time and location, grains with different composition reach the same value of $\zeta$ before any of the dynamical forces (drift and settling) separate them. We demonstrate in appendix~\ref{densvsgrowth}, that this is never the case.

In our simulations, we do not consider dust fragmentation. Fragmentation occurs when the relative velocity of the particles becomes greater than the critical velocity (the velocity between colliding particles which will likely lead to a fragmentation of the dust rather than sticking) \citep{2008ARA&A..46...21B}.   The dynamics of growing and fragmenting dust of single composition have been  studied in detail by \citet{2005A&A...434..971D,2010A&A...513A..79B,2012A&A...539A.148B,2015P&SS..116...48G} and more recently by \citet{2017MNRAS.467.1984G}. 
\citet{2005A&A...434..971D} and \citet{2010A&A...513A..79B,2012A&A...539A.148B} showed that a balance between growth and fragmentation can result in a steady distribution of different sizes in discs, especially in the inner disc zones. \citet{2017MNRAS.467.1984G} found that growing and fragmenting grains can result in a self-induced dust trap.

The critical velocity is a function of the chemical composition and the porosity of the dust  \citep{2014ApJ...783L..36Y}.   Preliminary tests for our disc model with the most recent theoretical values of 30-36  m\,s$^{-1}$ derived from \citet{2014ApJ...783L..36Y} for silicate aggregates made  of monomers of  $0.1~\mu \rm m$, show that this critical velocity is hardly reached in our large disc, and the effects of fragmentation are negligible within the considered evolutionary time. We assume  that no changes in the critical velocities occur with time, size of grains (always considered as grains made of monomers of  $0.1~\mu \rm m$), and species. However, critical velocities of different species, such as ice, silicates, iron and sulfides,  are still uncertain  \citep{2009ApJ...702.1490W,2009MNRAS.393.1584T,2013A&A...559A..62W,2014ApJ...783L..36Y} and they need to be assessed before implementation. 

For this reason, we decided not to include fragmentation at this stage. Fragmentation of grains will be investigated in a future work.

\subsection{Simulations}
\label{simulations}

In order to understand the effect of multiple intrinsic densities in shaping the chemical content of our disc, we  run a simulation with the dust distribution listed in Table~\ref{frac-abundance} with grain growth. As mentioned in the previous section, the initial grain size, $s_{0}$, common for all particles, is set to $10\mu \rm m$.

Similarly to \citet{2005A&A...443..185B} and \citet{2008A&A...487..265L}, the simulation starts at t=0 with a gas disc which is relaxed\footnote{ Same as in \citet{2005A&A...443..185B} and \citet{2008A&A...487..265L} the gaseous disc is evolved for a given time to let the pressure and artificial viscosity  smooth  the initial velocity field and the shape of the gaseous disc. Moreover, particles which, at any given time, move out the disc boundaries are removed from the calculation.} for $t\sim8000$~yr. At that time, the dust particles are injected in the system on top of the gas particles (see \citet{2005A&A...443..185B} for details). The system then evolves for a total of $t\sim1.6\times10^{5}$ years, after which the disc reaches a steady state: the majority of the dust particles decouples  from the gas and settles in the midplane, while the grain-growth becomes less efficient. This simulation is run in high resolution with a total of 400,000 SPH particles.

\section{Results}
\label{results}

 In this section we present the  resulting evolution of  dust particles in our disc. First we focus on the growth profile~(\ref{graingrowth}) and then on the vertical settling~(\ref{vs}) and radial drift~(\ref{rm}) of the dust particles.

\subsection{Grain growth}
\label{graingrowth}

\begin{figure*}
{\includegraphics[width=2\columnwidth]{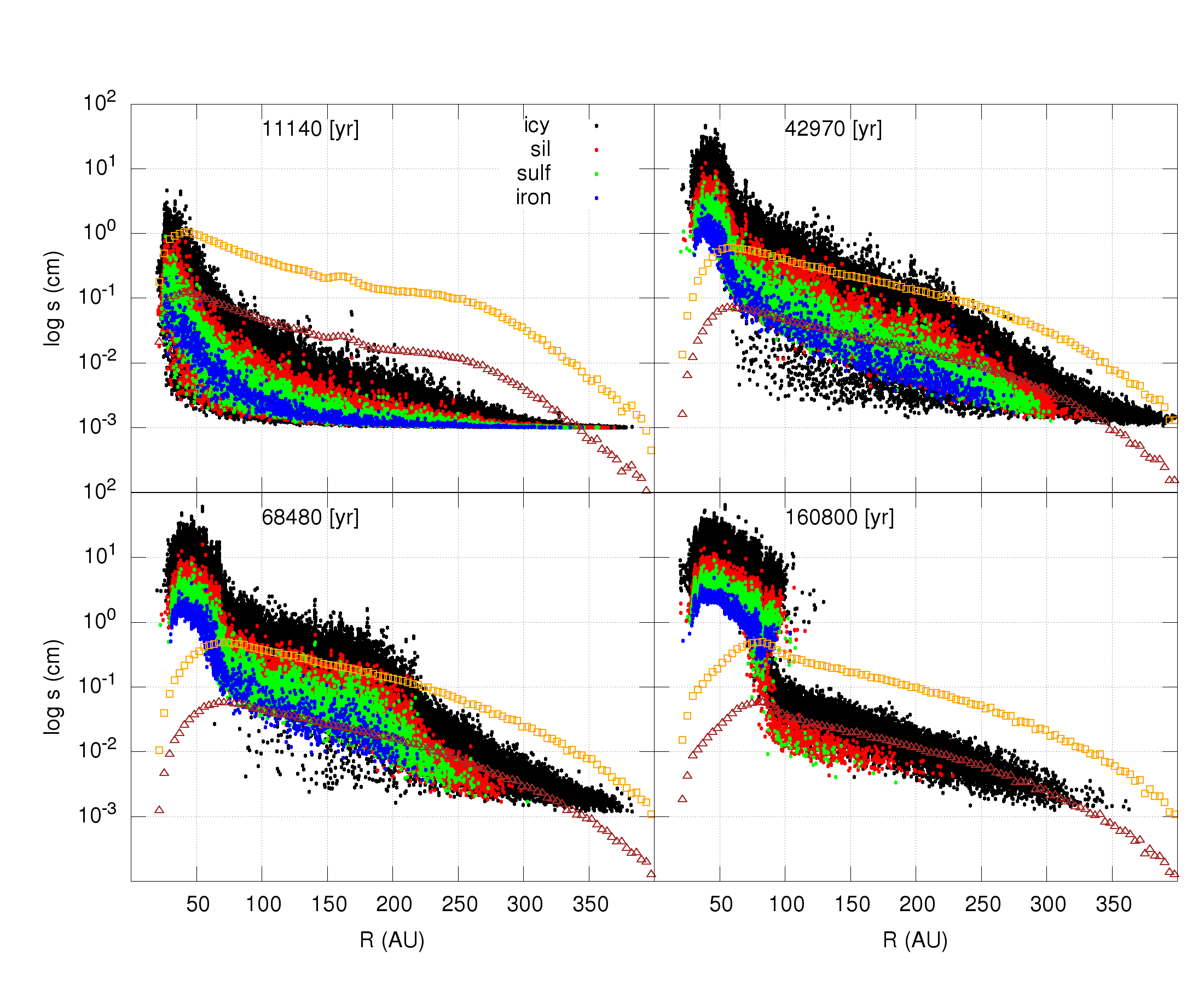}}\\
\caption{Radial size distribution for icy (black), which comprise water ice, volatile organics and refractory organics, silicates (red), sulfides (green) and iron (blue) particles at four evolutionary stages. Species are superimposed for ease of reading. The two curves represent the optimal drift size for water ice (orange with open squares) and iron (brown with open triangles), calculated using equation~\ref{optimalone2}. Grain growth {\bf proceeds} faster in the inner zones than in the outer zones. Moreover the optimal drift size is a function of the intrinsic density of the dust. Thus, denser grains will have a smaller optimal drift size compared to lighter grains. Since grains are growing, denser particles will start their radial drift at earlier stages in the outer disc, leading to a radial chemical sorting. \label{graingrow}}
\end{figure*}

Figure~\ref{graingrow} shows the evolution of  grain growth as a function of the distance from the star at four different evolutionary stages. The orange line with open squares and the brown line with open triangles represent respectively the $s_{\rm opt}^{\rm mid}$ for water ice and iron calculated following equation \ref{optimalone2}. Being function of the intrinsic density, the $s_{\rm opt}^{\rm mid}$ lines for silicates and sulfides, not shown in the figure, will lie in between.

 It can be seen that growth proceeds very rapidly in the inner disc  with particles reaching cm-sizes in a few thousand years. Similarly to \citet{2008A&A...487..265L}, our simulation shows that efficient growth is experienced by grains in the inner region and, as expected, we see that denser grains generally reach smaller sizes. This behaviour  results directly from the the prescription of growth derived by \citet{1997A&A...319.1007S}, with  $\mathrm{d}s/\mathrm{d}t\propto c_{\rm s}\rho_{\rm d}^{-1}$.
 
One may wonder why grains having reached their optimal drift size are not lost to the star but stay in the inner disk where they continue to grow, and whether this is a numerical artefact. In their SPH simulations, \citet{2004MNRAS.351..630L} found that the inner boundary conditions produced an outward erosion of the disc inner edge resulting in a gas pressure maximum slightly exterior to it, which would be capable of trapping drifting grains. However, the expectation that fast-drifting grains leave the disk before reaching large sizes comes from analytical \citep{1977MNRAS.180...57W} and numerical \citep{2008A&A...480..859B,2010A&A...513A..79B} work neglecting the drag back-reaction of dust on gas. On the contrary, several studies taking back-reaction into account \citep[see e.g.][]{2008A&A...487..265L,2010A&A...518A..16F,2015P&SS..116...48G,2016A&A...594A.105D} have shown that it slows down the drift of grains and, combined with their faster growth closer to the star, results in a pile up in the inner disc. This is also what happens in the simulation presented in this paper \citep[for a detailed discussion of the role of back-reaction, see][]{2017MNRAS.467.1984G}.

In \citet{2008A&A...487..265L}  the evolution of the radial size profile is driven by the rate of radial drift  caused by the optimal drift size of one species, ice. In our case, the evolution of the growth profile is not only driven by the size of the particles, but by the aerodynamic parameters ($\zeta$=$\rho s$), and by the optimal drift size proper to each species (see equations~\ref{optimalone} and~\ref{optimalone2}). The effects of grain growth on the evolution of the dust distribution will be discussed in detail in sections~\ref{rm} and~\ref{discussion}.

\subsection{Vertical settling}
\label{vs}

\begin{figure*}
\includegraphics[width=2.0\columnwidth]{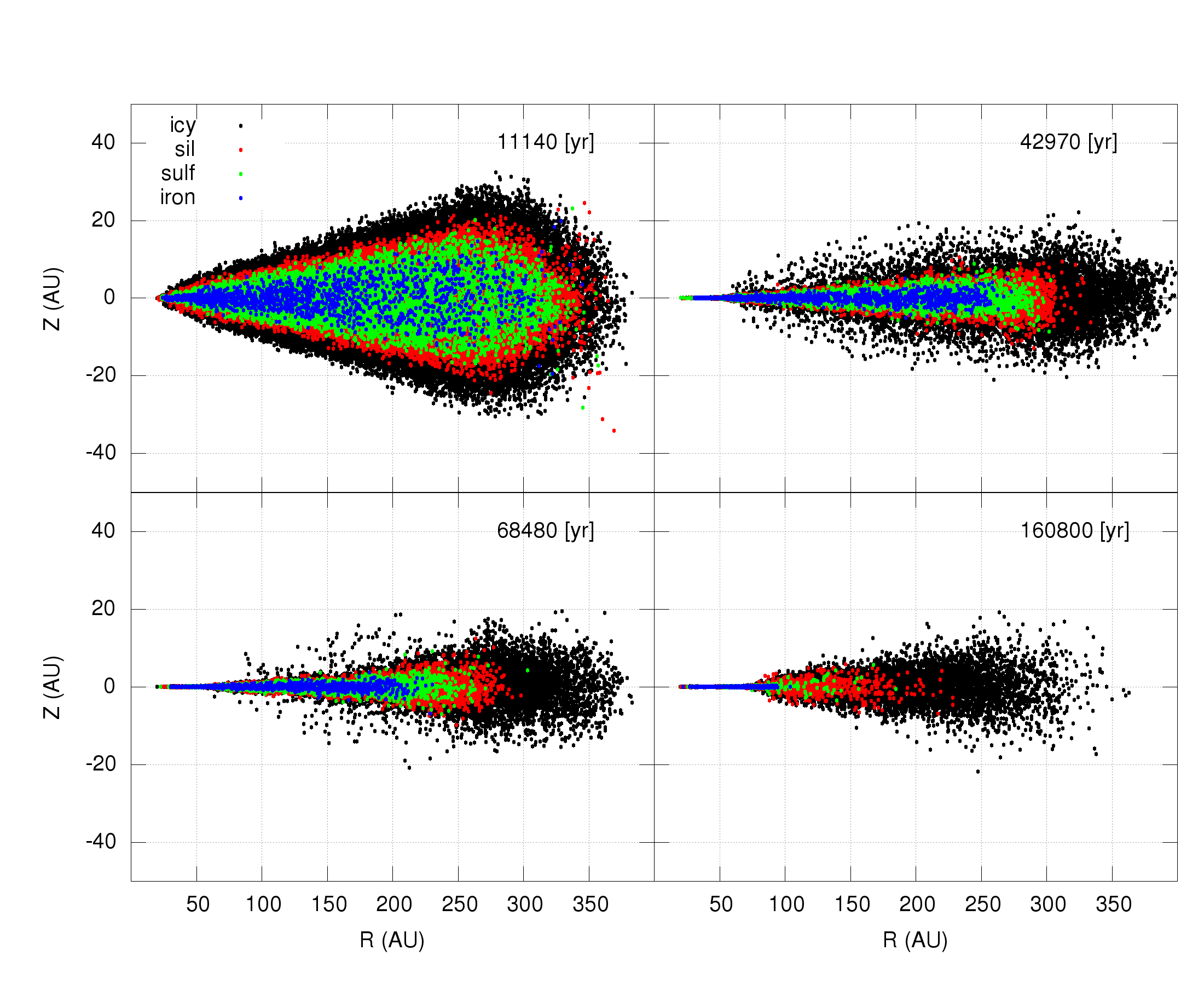} \\
\caption{Edge-on view of the spatial evolution for icy (black), silicate (red), sulfide (green), and iron (blue) grains at four evolutionary stages. Species are superimposed for ease of reading. A fast dust vertical settling is followed by an efficient dust radial drift. The effects of different dust chemistry in determining the rate of settling and drift are evident. \label{dustevo}}
\end{figure*}

Figure~\ref{dustevo} shows the distribution of the icy (water ice plus organics), silicates, sulfides and iron particles respectively, at four evolutionary stages for the disc seen edge-on: a fast vertical settling is followed by a very efficient radial drift of denser particles.  It is evident that the chemical characterization of the dust particles is driving the dynamics of the dust at different rates. 

\begin{figure}
\includegraphics[width=1.0\columnwidth]{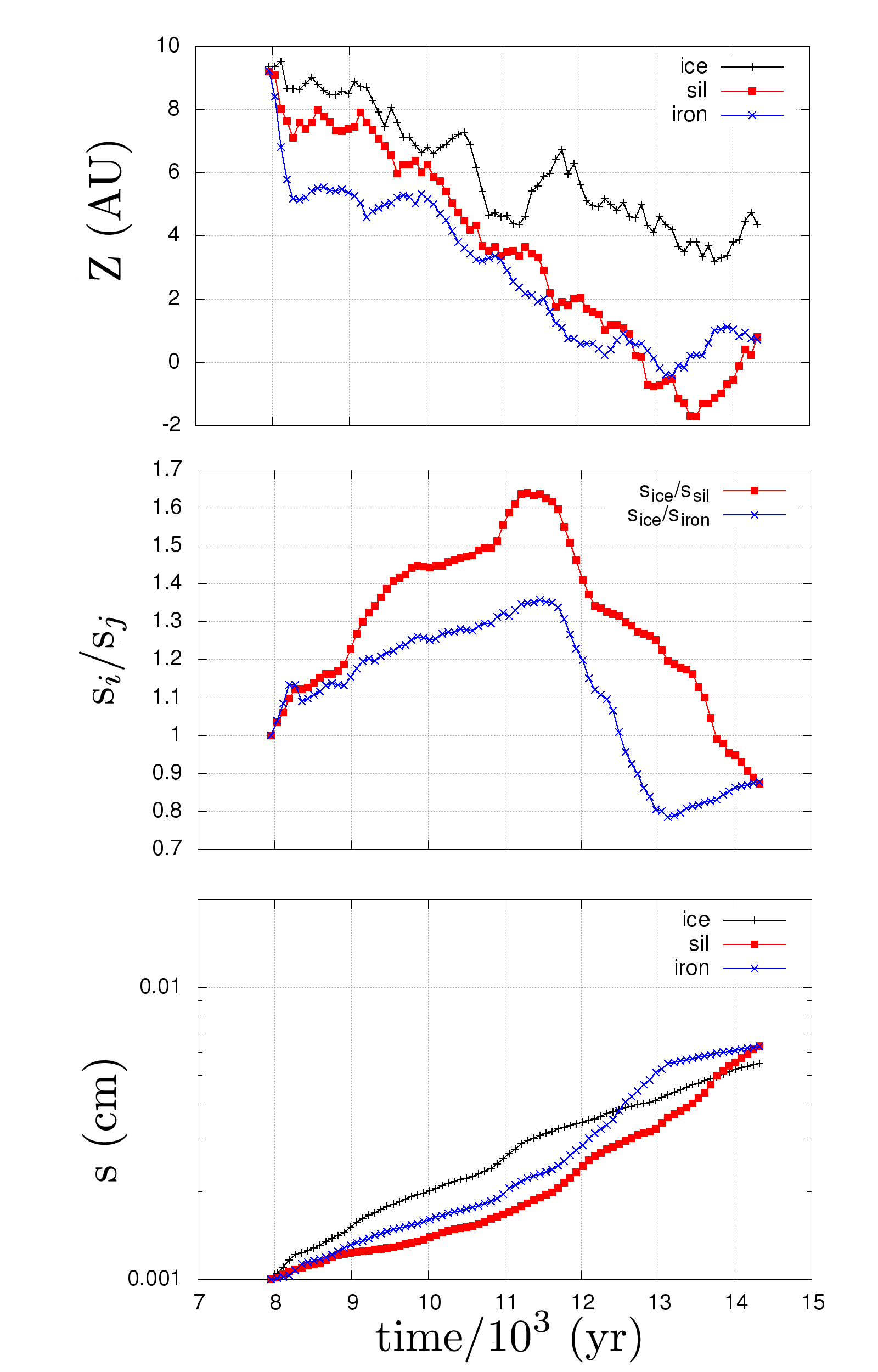}\\
\caption{Top: time evolution of the altitude of water ice (black), silicate (red), iron (blue) particles which lie at the same position (R$\sim$100~au, Z$\sim$9~au) at the beginning of the simulation. Middle: time evolution of  $s_{\rm ice}/s_{\rm sil}$ (red) and $s_{\rm ice}/s_{\rm iron}$ (blue) ratio. Bottom: size as a function of time for the three different particles. At the beginning of the simulation when the size of the particles is comparable, the vertical settling is density driven and will lead to a vertical chemical separation according to the intrinsic density of the grains. The curve of the $s_{\rm i}/s_{\rm j}$ ratios reaches a peak after 11 kyr and then decrease: ice particles grow bigger, thus the ratio increases. When the silicates and iron particles settle toward the midplane, they grow faster than ice, as they are in a denser environment, and thus the $s_{\rm i}/s_{\rm j}$ ratios starts to decrease. However, after a few thousand years we are already tracing particles which are far from each other and, thus, experiencing  different local conditions.\label{parttrack}}
\end{figure}

\begin{figure*}
{\includegraphics[width=2\columnwidth]{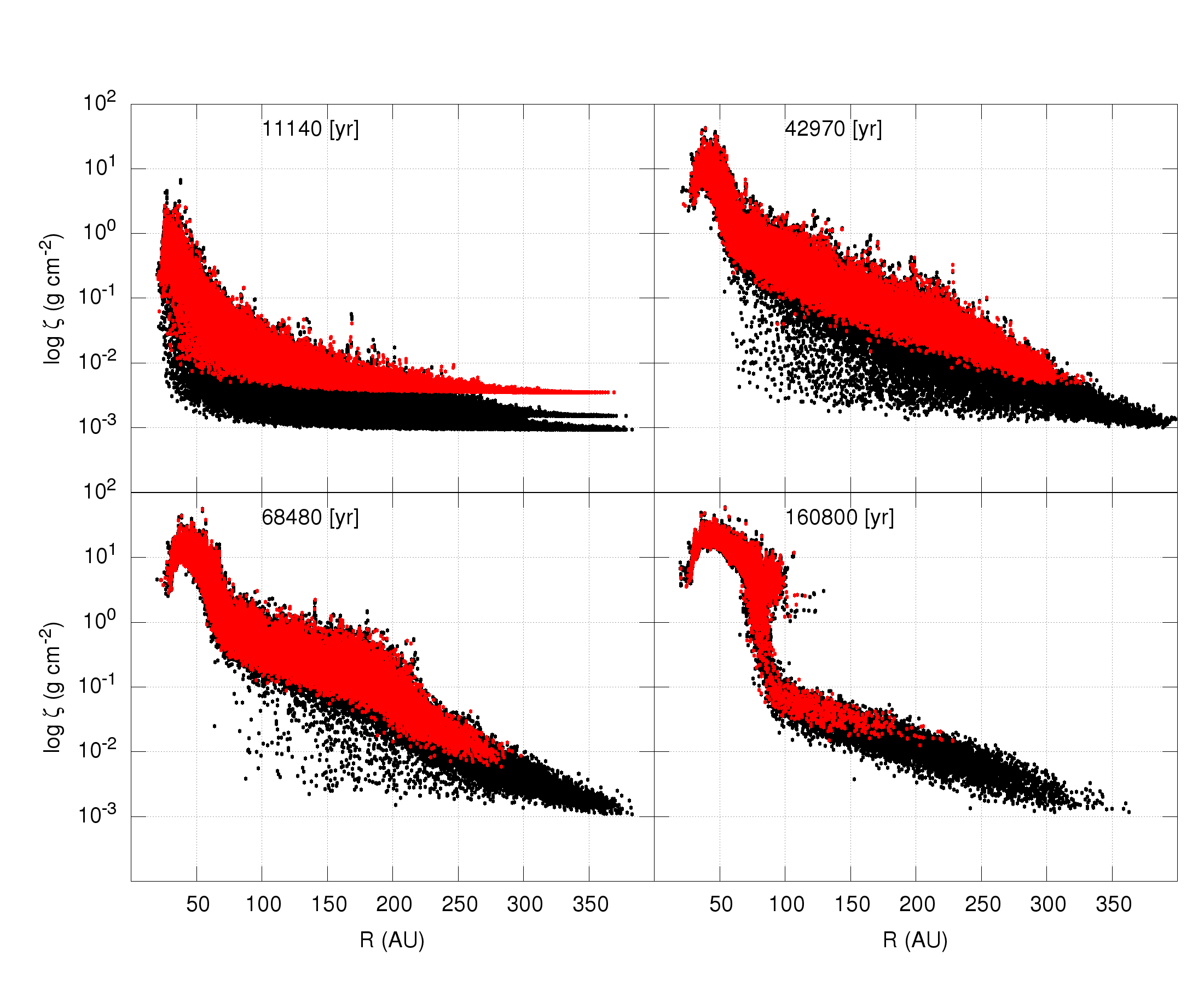}}
\caption{Aerodynamic parameter for icy (black) grains and silicates (red) grains at four evolutionary stages. The double tail for the icy grains is caused by the fact that we are considering all the three icy populations (water ice, volatile organics and refractory organic) which have intrinsic density $0.92\le \rho_{\rm d} \le 1.5$. When the size of the grains with different density is comparable, the heavier grains separate from the lighter grains and start to settle toward the midplane. As the lighter grains grow, their aerodynamic parameter increases and they start to behave as heavier grains and settle toward the midplane. After $t\sim$50~kyr, the vertical chemical separation of grains in the inner disc out to 80~AU is lost. \label{aeroicesilicates}}
\end{figure*}

In Fig.~\ref{parttrack}(top) we report  the altitude Z as a function of time  for three particles with different intrinsic density (water ice, silicate and iron) which are located  at the same position (R$\sim$100~au, Z$\sim$9~au) at the beginning of the simulation. In Fig.~\ref{parttrack}(middle) we report  the ratio $s_{\rm ice}/s_{\rm sil}$ (red) and $s_{\rm ice}/s_{\rm iron}$ (blue) as a function of  time, while  Fig.~\ref{parttrack}(bottom) shows the grain size as a function of time for the three considered species. In  Fig.~\ref{aeroicesilicates} we show the evolution of the aerodynamic parameter for icy particles (black) and silicate particles (red) at the same evolutionary stages as in Fig.~\ref{dustevo}. Figure~\ref{aeroicesilicates}  allows to analyse the global evolution of the aerodynamic parameter of all the selected particles.

At early stages of our simulation, the dust motion is dominated by  vertical settling which lasts $t\sim 40-50$~kyr. Looking at Fig.~\ref{dustevo}  we can distinguish two phases of vertical settling. First, the disc experiences a vertical chemical sorting with heavier particles populating a thin section of the disc. The disc assumes a chemically layered structure with icy particles occupying a larger vertical section. Moving from the surface of the disc toward the midplane, the icy disc becomes enriched in silicates, and then in sulfides. In the midplane, the dust will be iron-enriched.  This efficient vertical separation occurs over very short timescales. In the case reported in Fig.~\ref{parttrack}, particles reach their maximum vertical separation within a few hundred years. As the dust is injected at t$\sim$8~kyr, within t$\sim$3~kyr the disc is chemically vertically sorted in its large scale (see Fig.~\ref{dustevo}). 

This first phase of settling can be better explained by looking at the timescale of vertical settling for a dust grain in a disc, keeping in mind that at the beginning of the simulation grains have the same size ($\zeta_i$ differs only because of  their intrinsic density). From \citet{2004A&A...421.1075D},

\begin{equation}
\label{timesettling}
t_{\rm sett}= \frac{Z}{v_{\rm sett}}= \frac{4}{3}\frac{\sigma}{m}\frac{\rho_{\rm g} c_{\rm s}}{\Omega_{\rm k}^{2}},
\end{equation}
where $m$ is the mass of a spherical particle and $\sigma$ its cross section. Given that $(\sigma/m)\propto (\rho_{\rm d}s)^{-1}= \zeta^{-1}$, equation~\ref{timesettling} states that, for given conditions,  (${\rho_{\rm g} c_{\rm s}}/{\Omega_{\rm k}^{2}}$), and distance to cover, $Z$, larger-heavier particles will have smaller $t_{\rm sett}$ or will settle at higher speed than smaller-lighter particles. In this case, two grains made of different material, $i$ and $j$, but with the same size would have $(t_{\rm sett}^{i}/t_{\rm sett}^{j})\propto (\rho_{\rm j}/\rho_{\rm i})$. As an example the $t_{\rm sett}^{ice}$ is 8.55=(7.87/0.92) times longer than $t_{\rm sett}^{iron}$, and in order to erase the effect of density ice particles should grow 8.55 times larger than the iron particles before they  separate. Figure~\ref{parttrack}~(middle) and Fig.~\ref{parttrack}~(bottom) show that during the first phase of vertical settling growth is not efficient to counterbalance the strong effect brought by the intrinsic density of grains (in the first couple of kyr of our simulation the $s_{\rm ice}/s_{\rm iron}$ ratio is always in the order of 1-1.2. Thus, when the size of two particles is comparable, which is the case in the early evolution, the vertical settling is density-driven.

A second phase of vertical settling starts immediately in the inner disc and, at later times, in the outer disc:  the icy particles settle toward the midplane, remixing with the silicate-iron-rich dust. This second stage of vertical settling can be explained by the evolution of the aerodynamic parameter for  single species. In Fig.~\ref{aeroicesilicates} the icy particles for which  $\zeta_{ \rm icy}<\zeta_{\rm sil}$, are the particles which populate the icy contour of the disc in Fig.~\ref{dustevo}. As grain growth proceeds, $\zeta$ of the lighter material on the surface of the disk increases and becomes comparable to the values which characterised the denser material at earlier stages. We have a superimposition of the  $\zeta$ values with  $\zeta_{\rm icy}\sim\zeta_{\rm sil}$ and the vertical settling of the lighter material becomes more efficient. This phase of settling is, thus, size driven.

 The introduction of lighter material into the midplane is more efficient in the inner zone of the disc (see Fig.~\ref{dustevo}).  At the end of our simulation and in the inner $\sim80$~au, all the icy particles are introduced in the denser midplane. In the outer part of the disc a vertical chemical sorting can still be noticed. This behaviour is due to the lower growth rate in the outer disc (see Fig.~\ref{graingrow}, and \citet{2008A&A...487..265L}), where smaller and lighter icy grains will remain coupled to the gas longer.

\subsection{Radial drift}
\label{rm}

\begin{figure}
{\includegraphics[width=1.0\columnwidth]{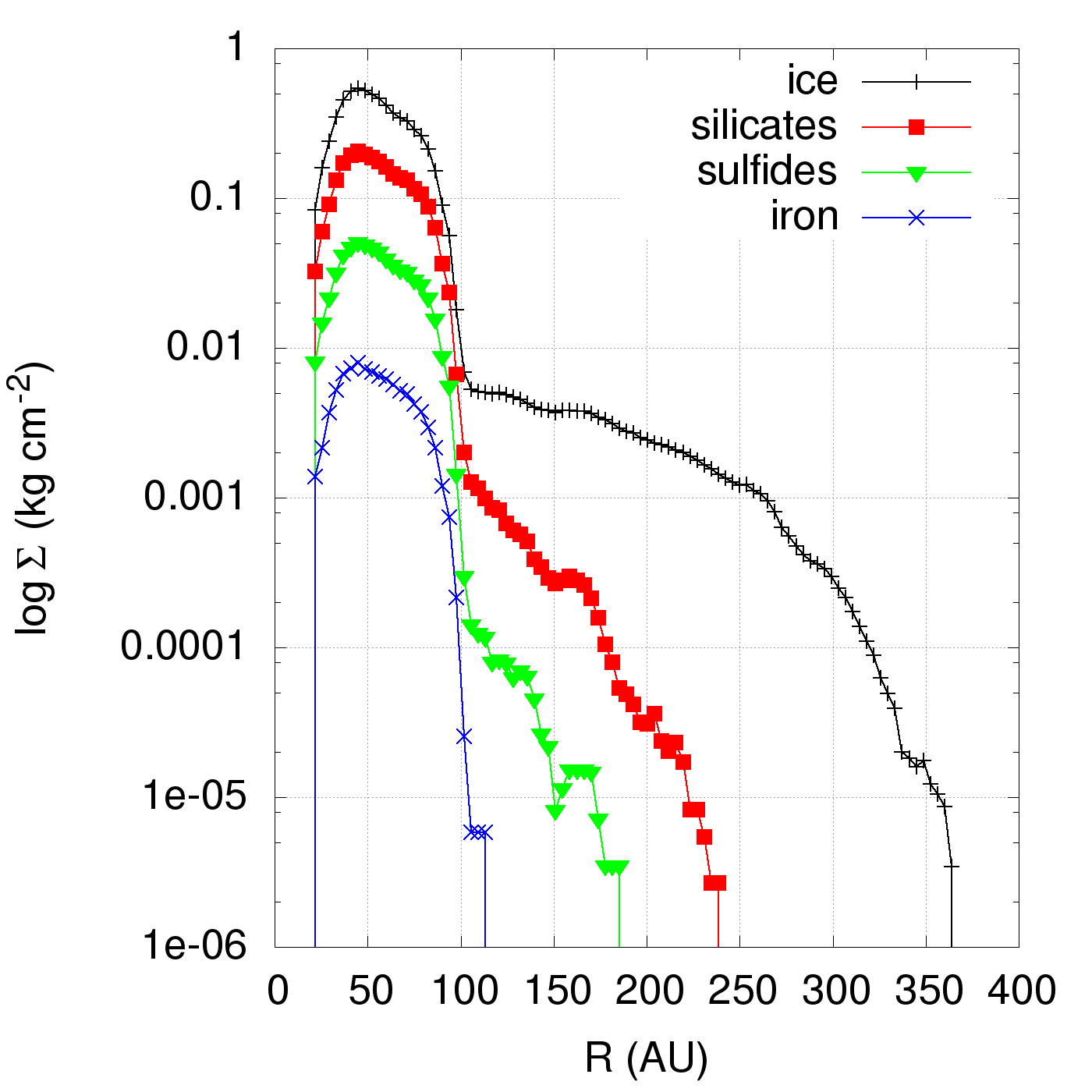}} 
\caption{Surface density of water-ice, silicates, sulfides and iron particles at the end of the simulation. The chemical radial sorting is evident. The differences in the values of the surface densities are due to the different initial amount which characterizes each species. \label{surfdust}}
\end{figure}

After the first $\sim40-50$~kyr, the radial drift of the dust starts to become extremely efficient for the denser grains (see Fig.~\ref{dustevo}). We see a radial sorting of chemical species with time. The outer disc becomes depleted in heavy particles in t$\sim$50~kyr.  Figure~\ref{surfdust} shows the surface density of different species at the end of the simulation: the disc outside $\sim$110~au is iron-depleted, outside $\sim$200~au sulfide-depleted and outside 250~au silicate-depleted. The regions beyond 250~au are ice-rich. The different maximum values of the surface densities  are due to the differences in the initial amount which characterizes  each species.

As reported in the introduction, the dust radial drift is driven by the optimal drift size proper to each dust particle (see equations.~\ref{optimalone} and~\ref{optimalone2}).
In Fig.~\ref{graingrow} we reported the evolution of the radial profile of $s_{\rm opt}^{\rm mid}$ for water ice and iron at different evolutionary stages. Iron particles have a smaller optimal drift size than water ice particles. Looking at Fig.~\ref{graingrow} it can be seen that the smaller optimal drift size for iron particle is leading to a radial chemical sorting of the dust. Similar to iron, sulfides and silicates follow the radial drift dictated by their $s_{\rm opt}^{\rm mid}$.  Not all the water-ice particles in the outer disc reach their optimal drift size by the end of the simulation, making the radial drift of ice overall less efficient.

\begin{figure}
{\includegraphics[width=1.0\columnwidth]{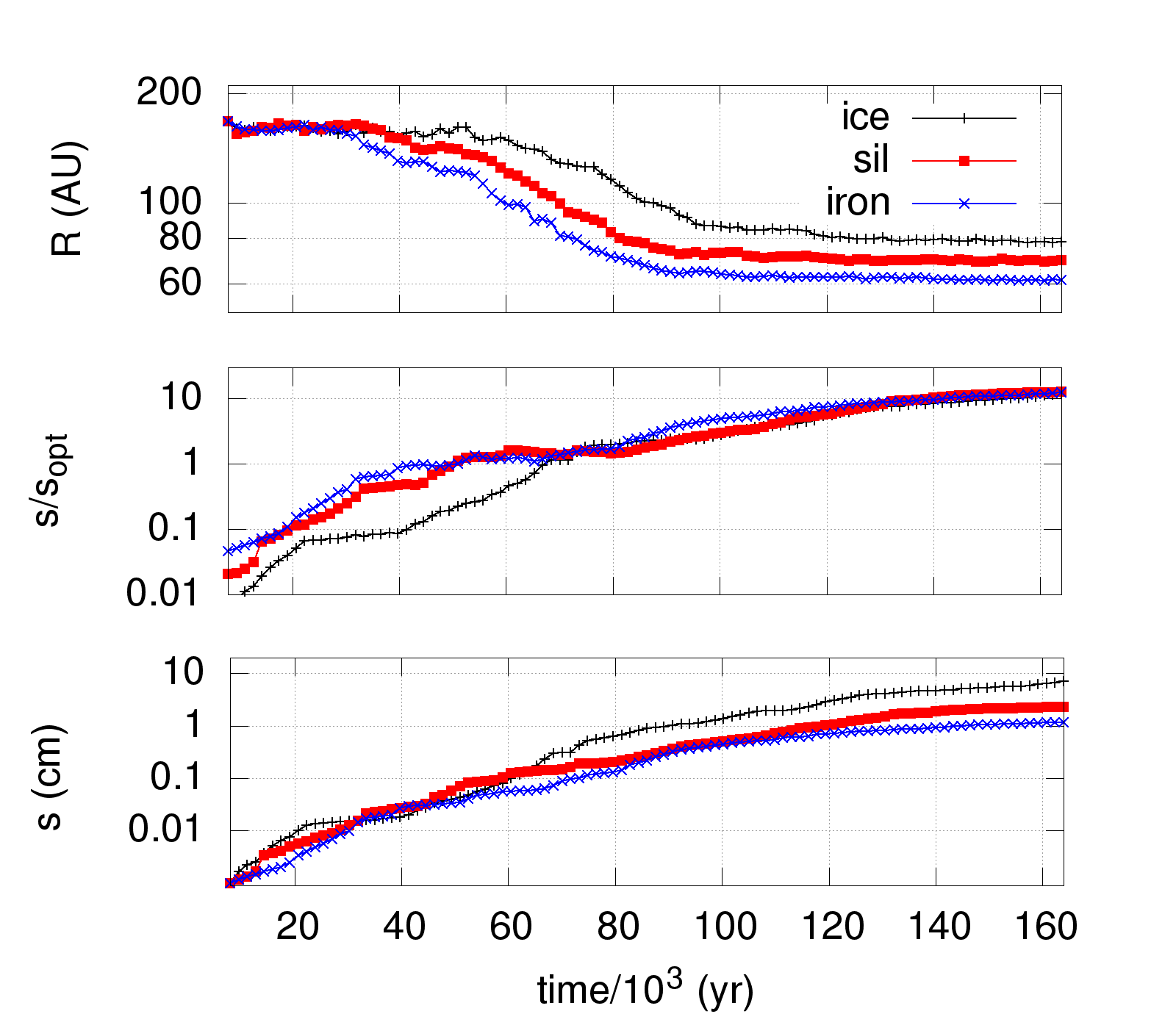}} \\
\caption{Top: radial drift for water ice, silicates and iron particles as a function of time. $s/s_{\rm opt}$ (middle) and s (bottom) as a function of time for the same particles. Denser particles have smaller optimal drift size. As particles grow, iron particles reach their optimal drift size before lighter particles and thus, start to drift inward before the lighter particles.\label{trackgrowth}}
\end{figure}

Figure~\ref{trackgrowth} tracks the radial drift (top),  $s/s_{\rm opt}$ (middle), with $s_{\rm opt}$ calculated using equation~\ref{optimalone}, and the size  (bottom) as a function of time for three particles with different instrinsic density (water-ice, silicate and iron) which are located at the same position at the beginning of the simulation. As the growth  proceeds, the iron particle reaches the $s/s_{\rm opt}=1$  point before the lighter particles and starts to drift inward more efficiently. Lighter particles keep growing in place before reaching their optimal size and starting to drift toward the inner regions of the disc.

\section{Discussion}
\label{discussion}

Our results are in good agreement with those found by \citet{2008A&A...487..265L}, \citet{2014MNRAS.437.3037L}, \citet{2014MNRAS.437.3025L},  and \citet{2014MNRAS.437.3055L}, who studied the dynamics of growing but not fragmenting grains illustrated in section~\ref{growthpresc}. Furthermore, we confirm that, (i) a very efficient vertical chemical sorting of the dust starts at the beginning of the simulation, and (ii) the density driven vertical settling which produces the initial chemical sorting is the first dynamical mechanism expected when the size of the grains is comparable. Moreover, the grain growth within our disc model is not efficient enough to counterbalance this process. 

Our results show that, as the intrinsic density of the growing grains determines the optimal drift size, the radial drift of different species would produce a radial chemical sorting of the dust. As a consequence of the vertical settling and radial drift, the disc will lose its initial chemical homogeneity, with inhomogeneities occuring from the beginning of the simulation. These results could bring dramatic consequences in the composition of the building block of meteorites parent bodies and planetesimals. The chemical and dynamical evolution of the dust in our disc is thus the main focus of this section. We will discuss the physical properties of the particles in section \ref{sidenssorting} and the evolution of the chemical content in our disc in section \ref{chemsorting}.

\subsection{Size and density sorting}
\label{sidenssorting}

\begin{figure}
{\includegraphics[width=1\columnwidth]{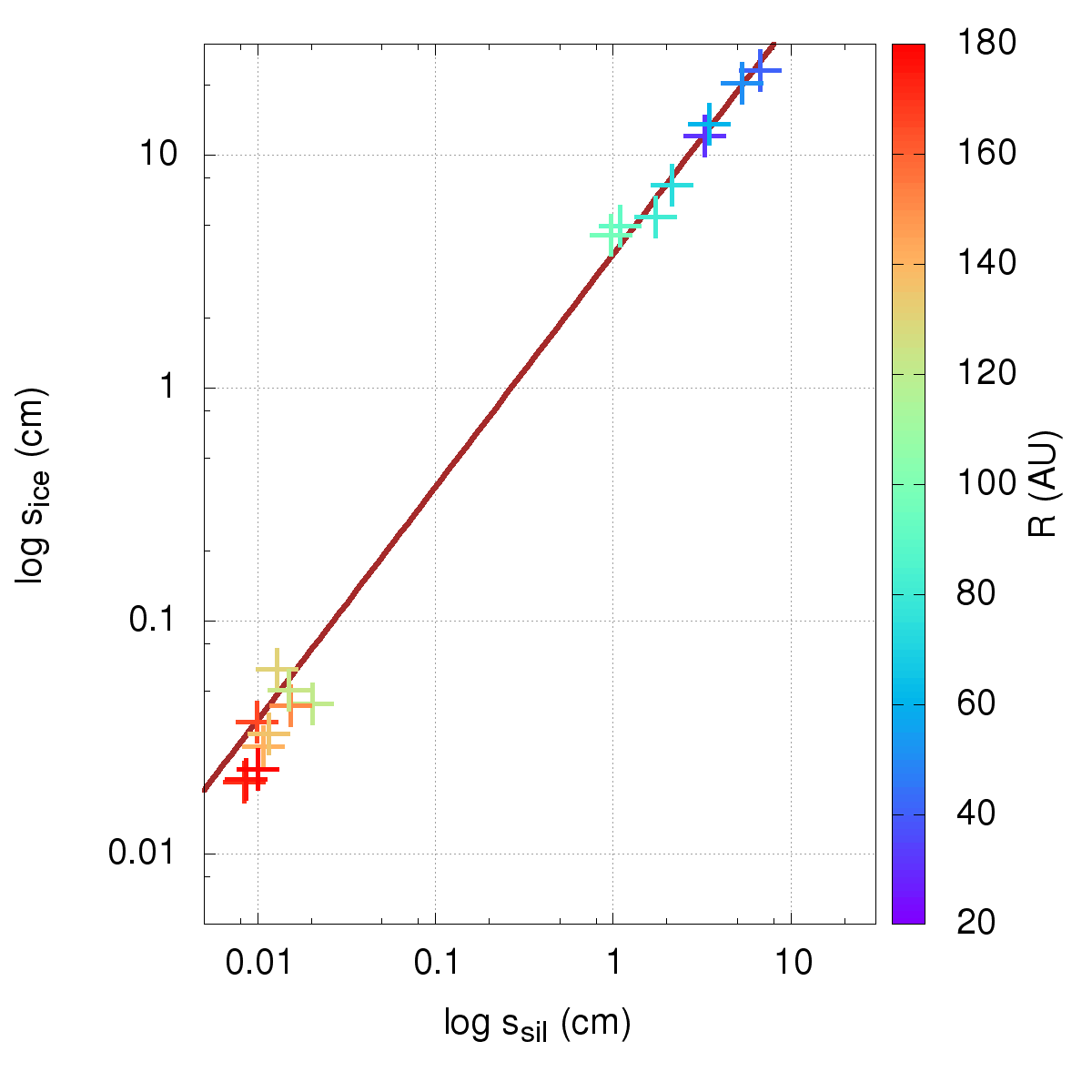}}
\caption{Average size of silicates versus average size of ice particles at different distances from the protostar (color). The brown line represents the theoretical size of ice grains predicted using equation.~\ref{sizesorted}. The dust in the inner disc ($R<100$~au) is aerodynamically sorted with a correlation coefficient of 0.99. This sorting is less evident beyond 100~au, where the predicted thrend is not fully satisfied with a correlation coefficient of 0.64.\label{icesilicatesorted}}
\end{figure}

Using our set of diagnostic we are able to trace all the properties of every SPH particle at any given time. At the end of the simulation, we randomly sampled groups of water ice and silicates particles which are located at the same coordinates in the midplane and record their size and intrinsic density. We replicate this measurement at different radial distances from the central star. We chose ice and silicates particles as they are still present at large radial distance at the end of the simulation. In Fig.~\ref{icesilicatesorted} we thus plot the average size of water-ice particles versus the average size of silicate particles  at different radial distances from the star (colours). The brown line represents the theoretical size of water-ice particles  predicted by the following equation
\begin{equation}
s_{\rm ice} = s_{\rm sil} \frac{\rho_{\rm sil}}{\rho_{\rm ice}} ,\
\label{sizesorted}
\end{equation}
i.e. the predicted size of water-ice particles if perfect size-density sorting with silicates particles occurs ($\zeta_{\rm ice}=\zeta_{\rm sil}=(s\rho)_{ \rm ice}=(s\rho)_{ \rm sil}$). 

The returned correlation coefficient between the average size of silicates and average size of the ice particles for all the data points within $R<100$~au in Fig.~\ref{icesilicatesorted} is $R_{\rm c}$=0.99, while for the points laying at $R\ge100$~AU the correlation coefficient is $R_{\rm c}$=0.64. The inner midplane of the disc is generally aerodynamically (size-density) sorted. This is less evident in the outer part of the disc, where the lower growth rate and the lower gas density causes the aerodynamic sorting to occur more slowly. Thus, in case of pure growth  particles tend to naturally sort in the disc according to their aerodynamic parameter. 

Our finding confirms that aerodynamic sorting can be an active mechanism which characterizes the disc at very large scales and suggest that the sorting trend observed in chondrites could thus  have occurred not only in a local environment but also at disc scales. Since the aerodynamic sorting in chondrites could have occurred between chondrules and metal-troilite grains within the same chondrite group \citep{1998LPI....29.1457B,2015ChEG...75..419F}, we investigate the aerodynamic sorting between different species in more detail in section~\ref{when}.

\subsection{Disc chemical and dynamical evolution}
\label{chemsorting}

As discussed in section~\ref{results}, vertical settling and radial migration continuously change the distribution of dust particles in different regions of the disc.

\begin{figure*}
{\includegraphics[width=0.65\columnwidth]{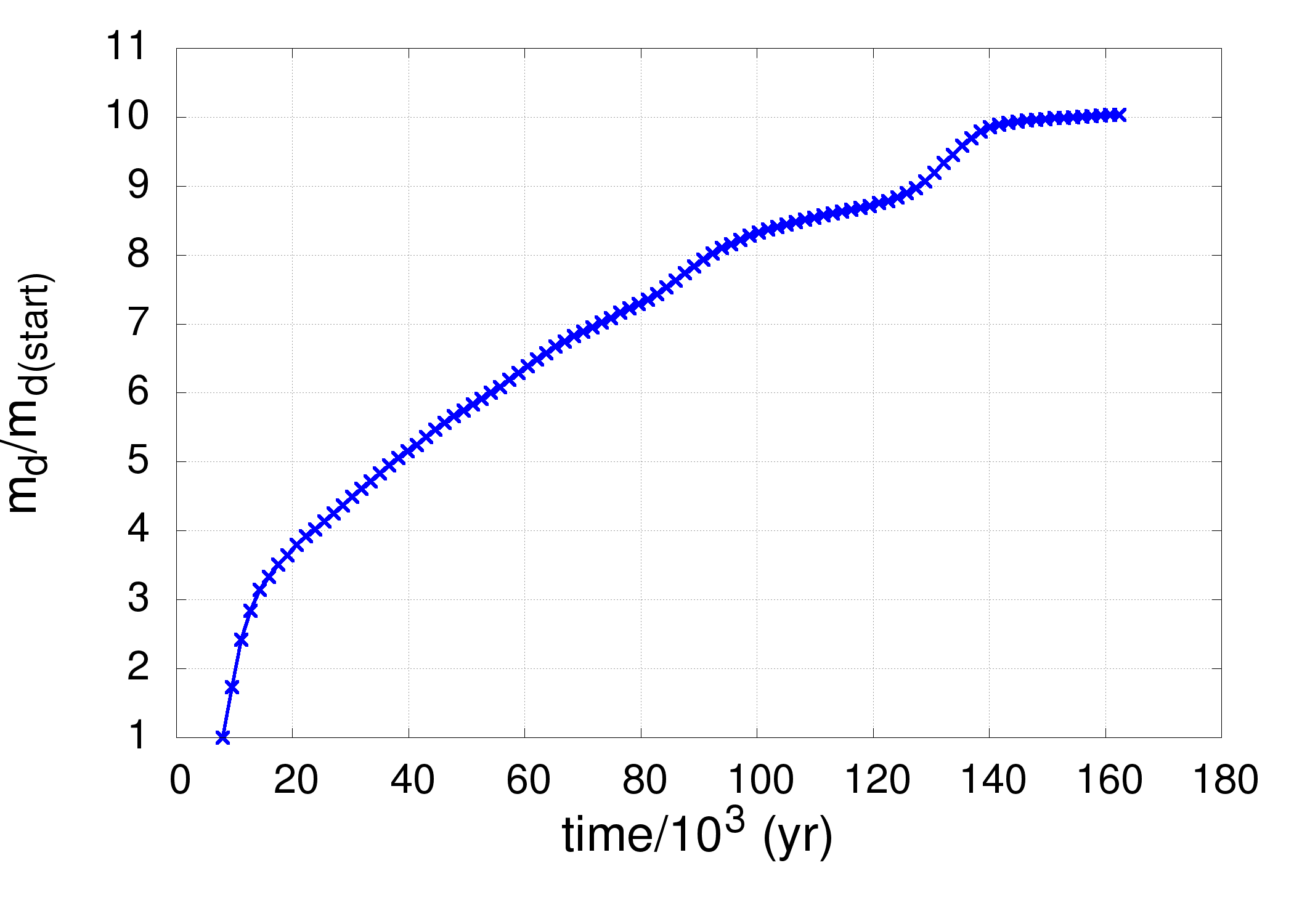}}
{\includegraphics[width=0.65\columnwidth]{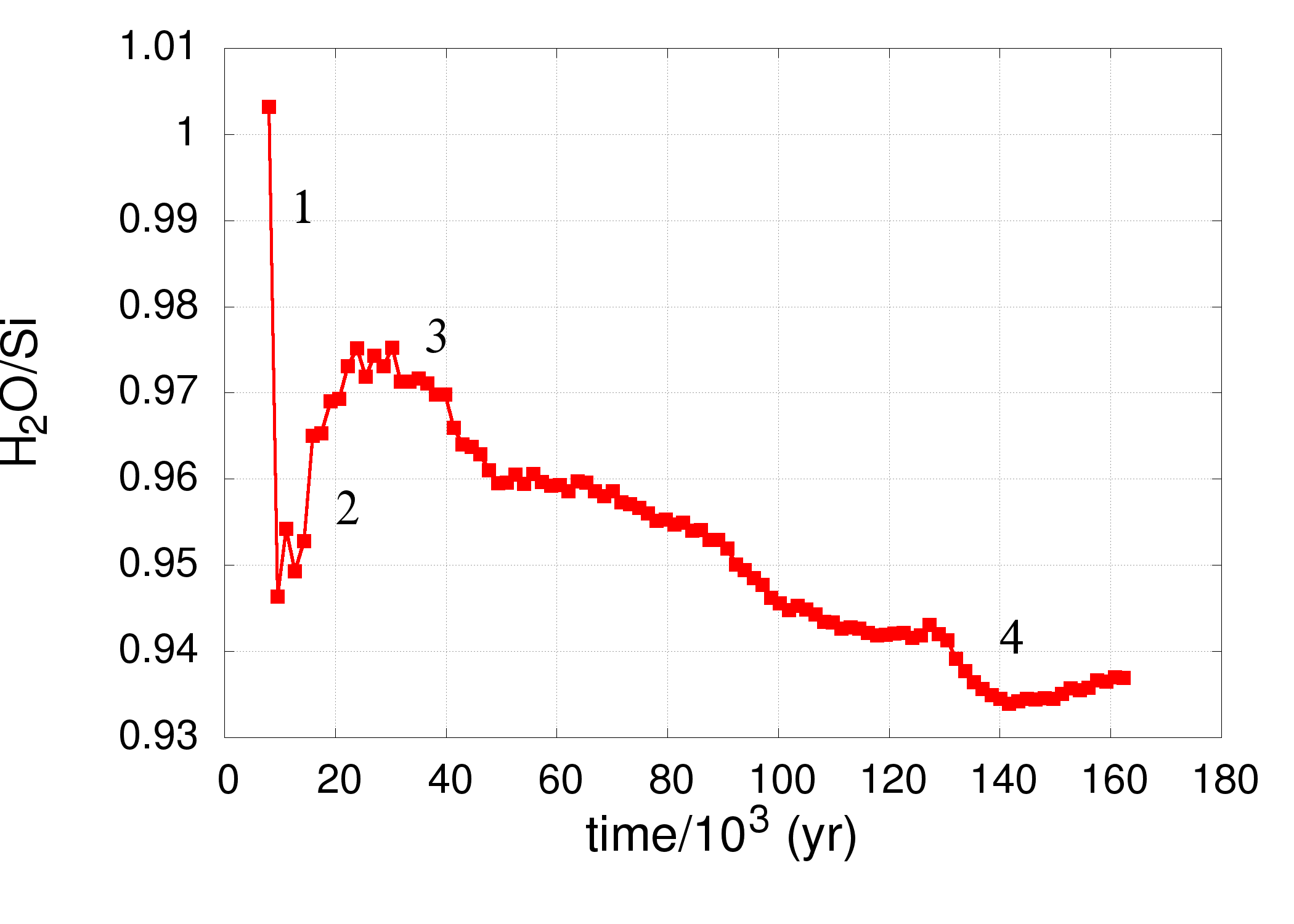}} 
{\includegraphics[width=0.65\columnwidth]{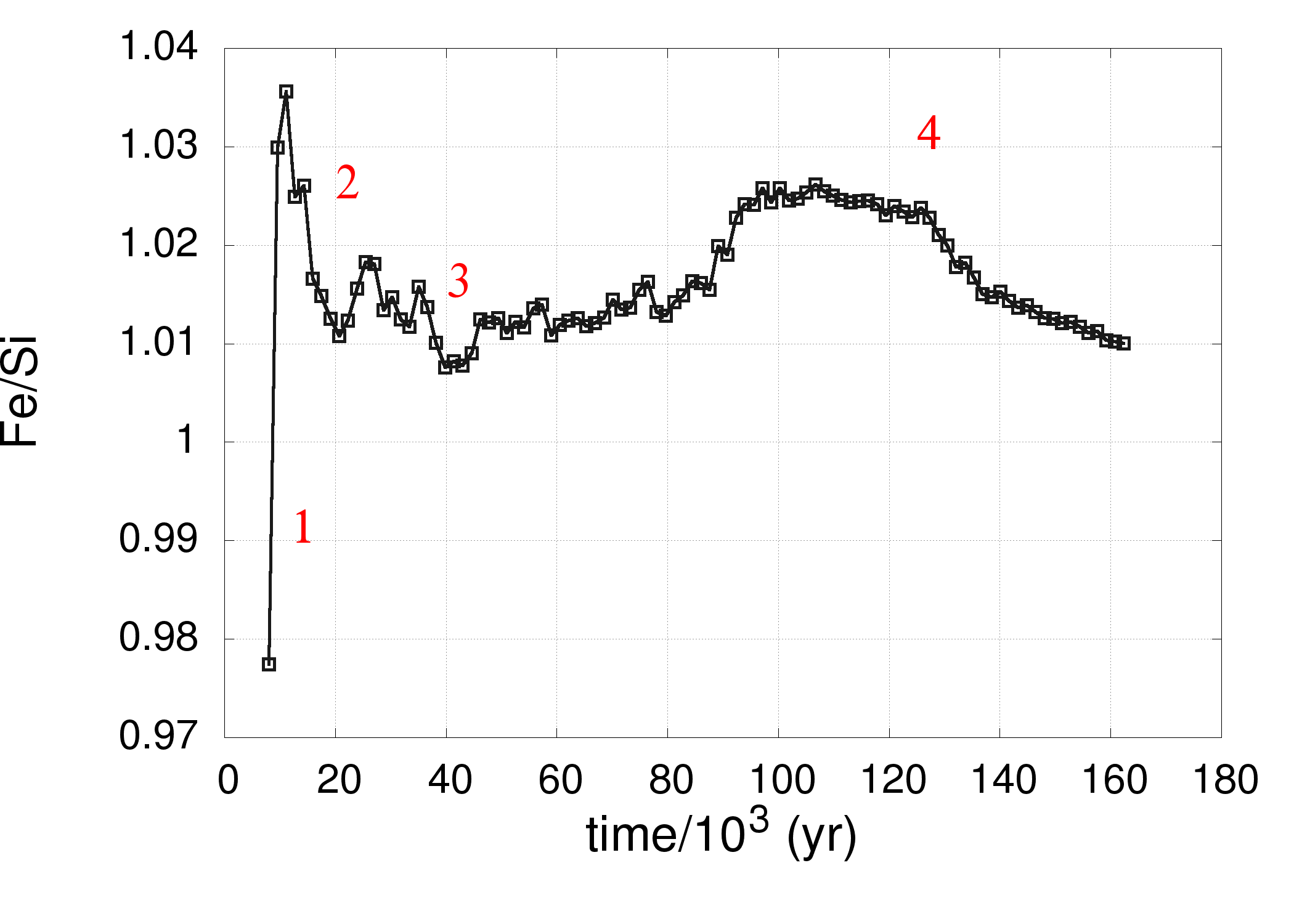}} \\
{\includegraphics[width=0.65\columnwidth]{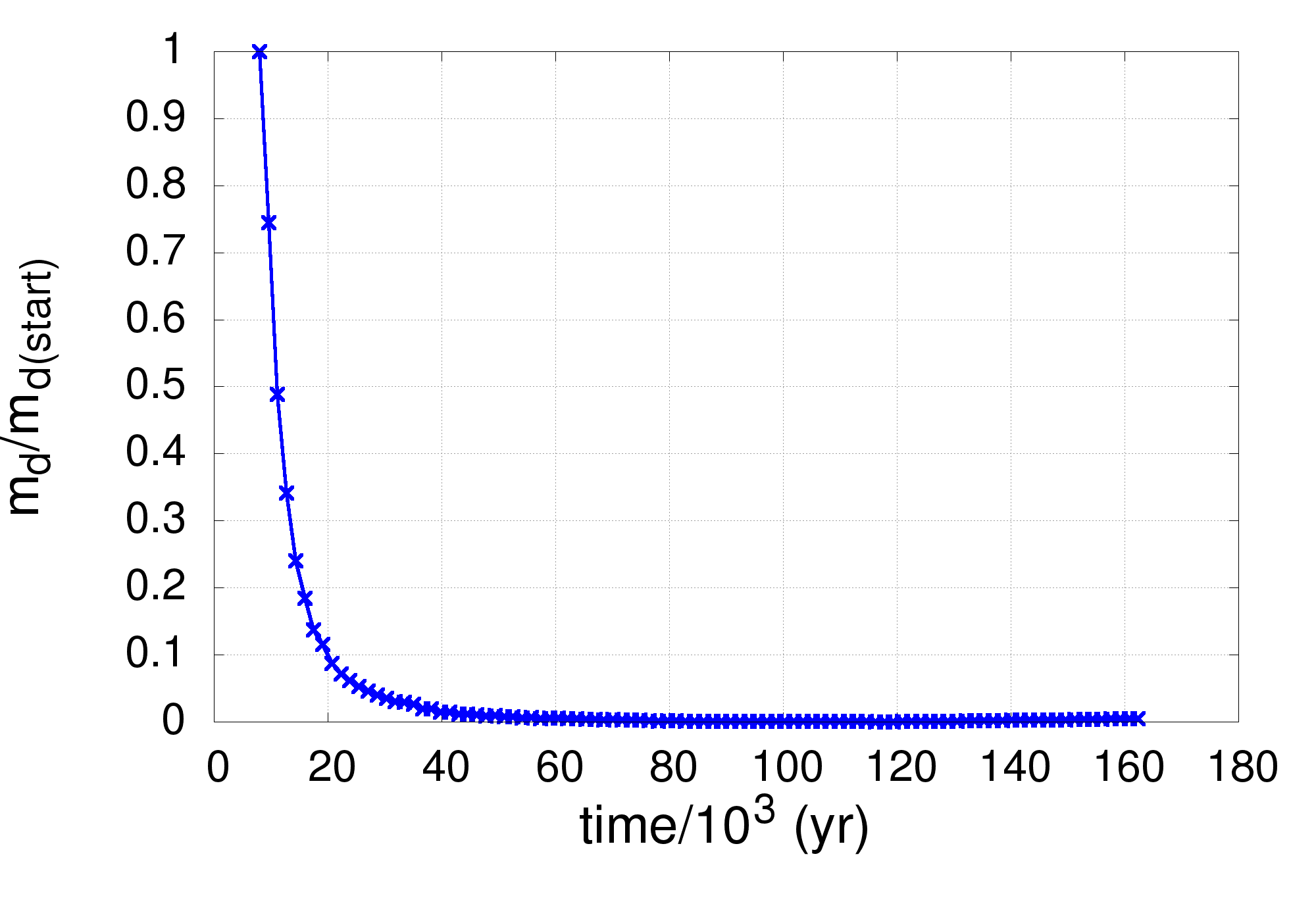}}
{\includegraphics[width=0.65\columnwidth]{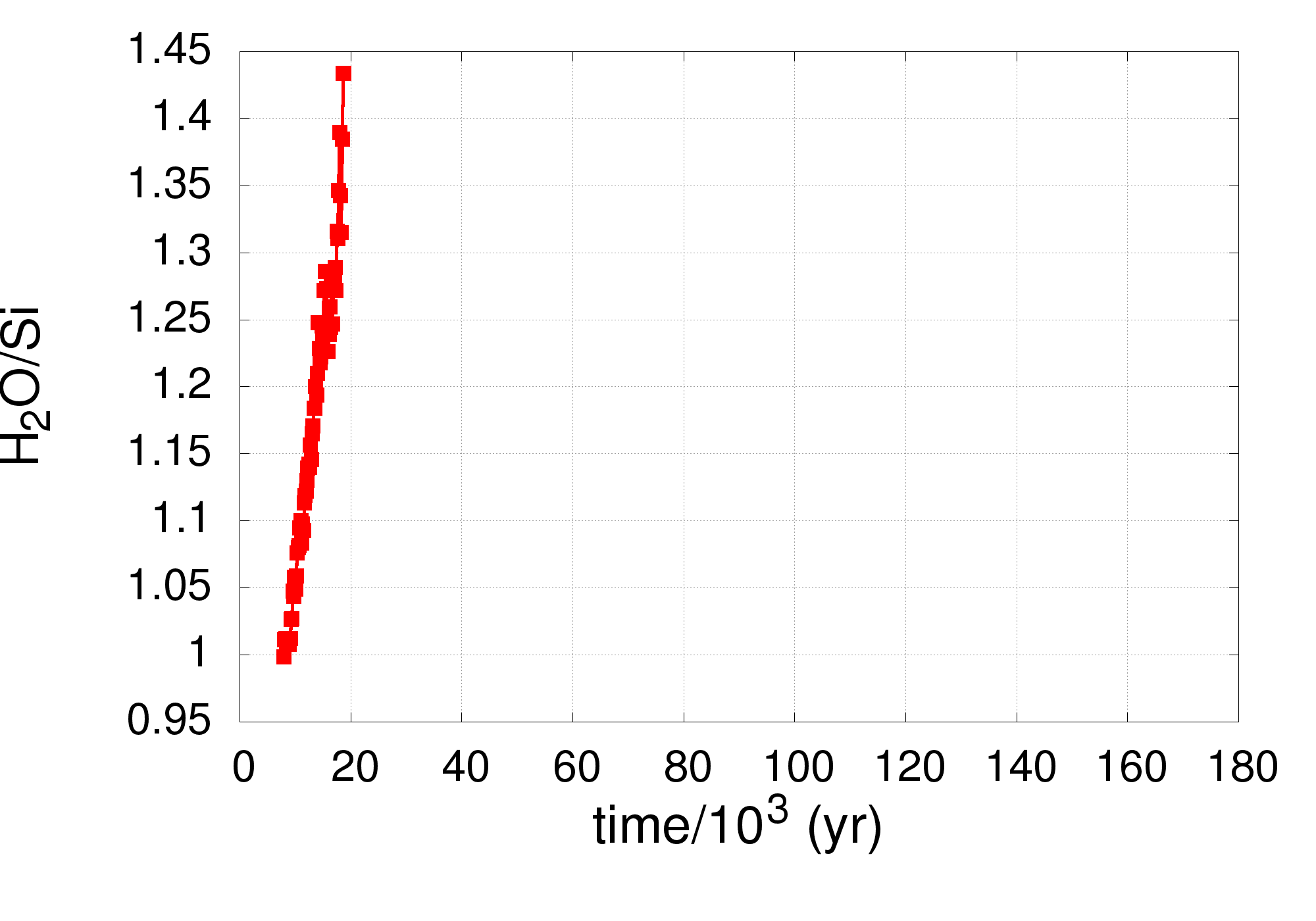}} 
{\includegraphics[width=0.65\columnwidth]{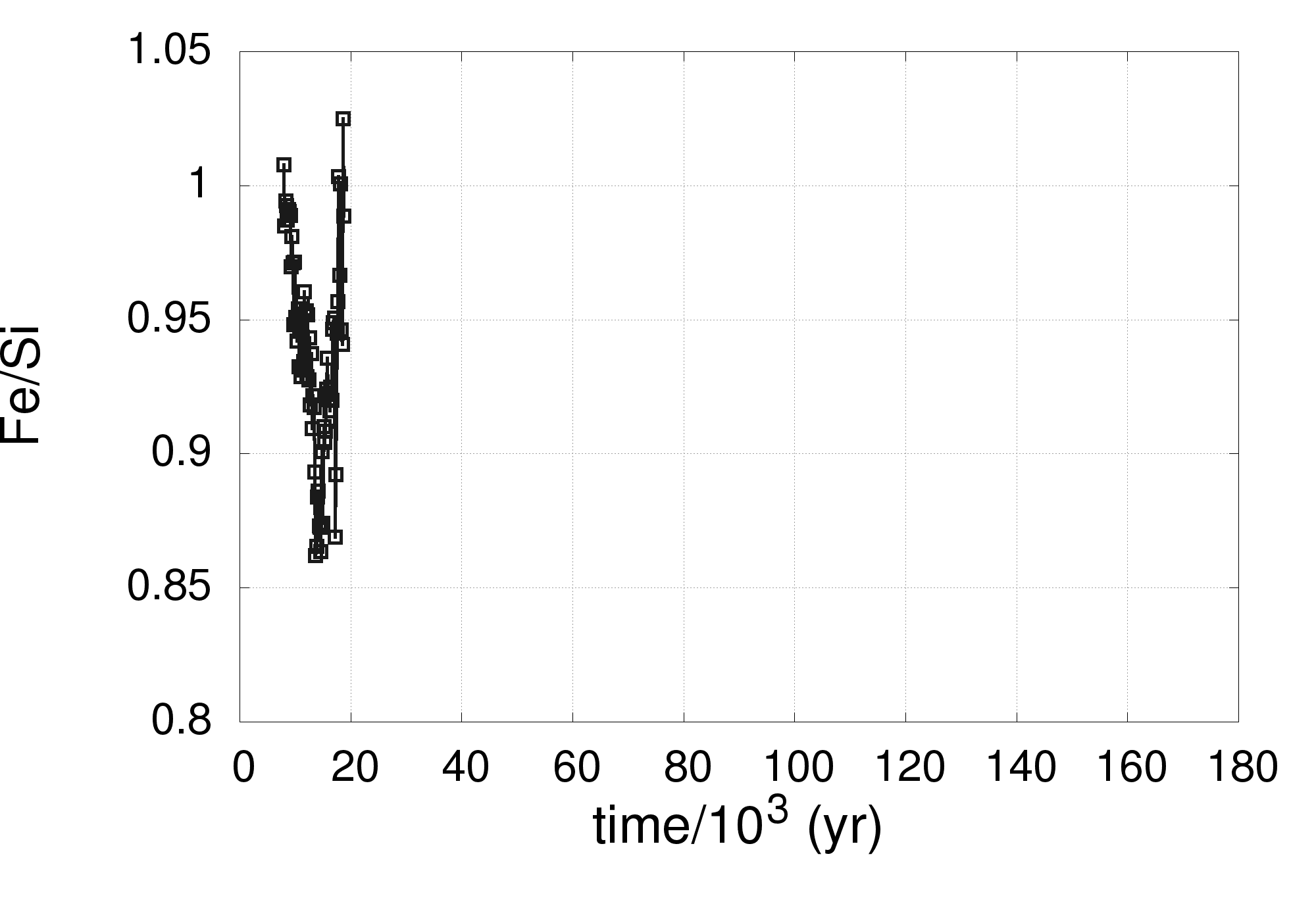}} \\
\caption{Top row: time evolution of the dust mass content compared to the initial mass present in the midplane, $-1<Z\rm{(AU)}<1$, where $20<R\rm{(AU)}<100$ (left),  \ce{H2O}/\ce{Si} (centre), and  \ce{Fe}/\ce{Si} (right) ratios in the considered zone. The changes in the chemical ratios brought by the dust dynamics are evident: (1) the density-driven vertical settling, (2) the size-driven vertical settling, (3) the efficient radial drift of denser dust, (4) the radial drift of the residual lighter dust. Bottom row: same as top row for the disc surface where $\lvert Z\rm{(au)}\rvert>1$.  The chemical ratios of the disc surface are plotted until $m_{\rm d}/m_{\rm d(start)}$ reaches 10\% of the initial mass.\label{100}}
\end{figure*}

\begin{figure*}
{\includegraphics[width=0.65\columnwidth]{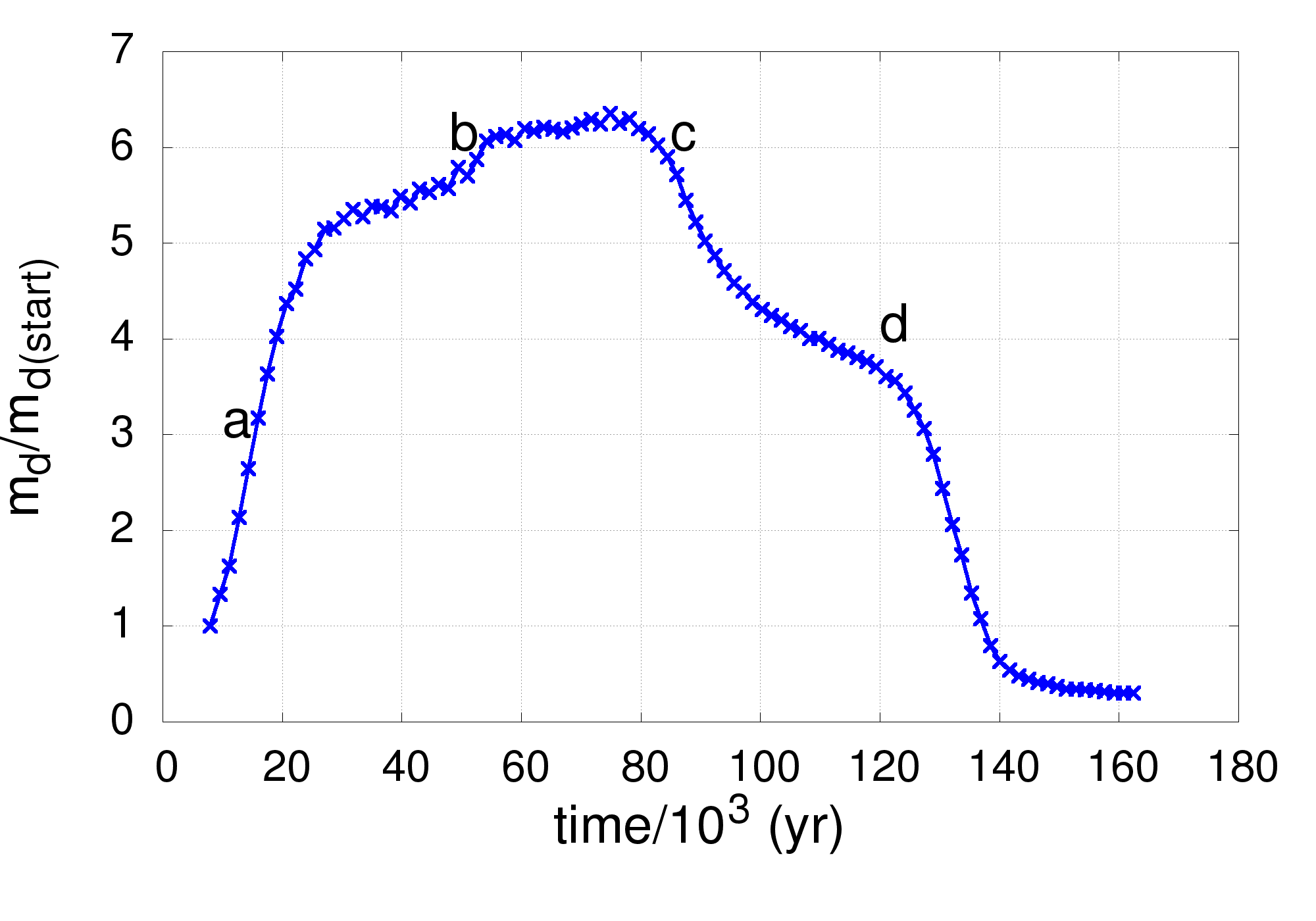}}
{\includegraphics[width=0.65\columnwidth]{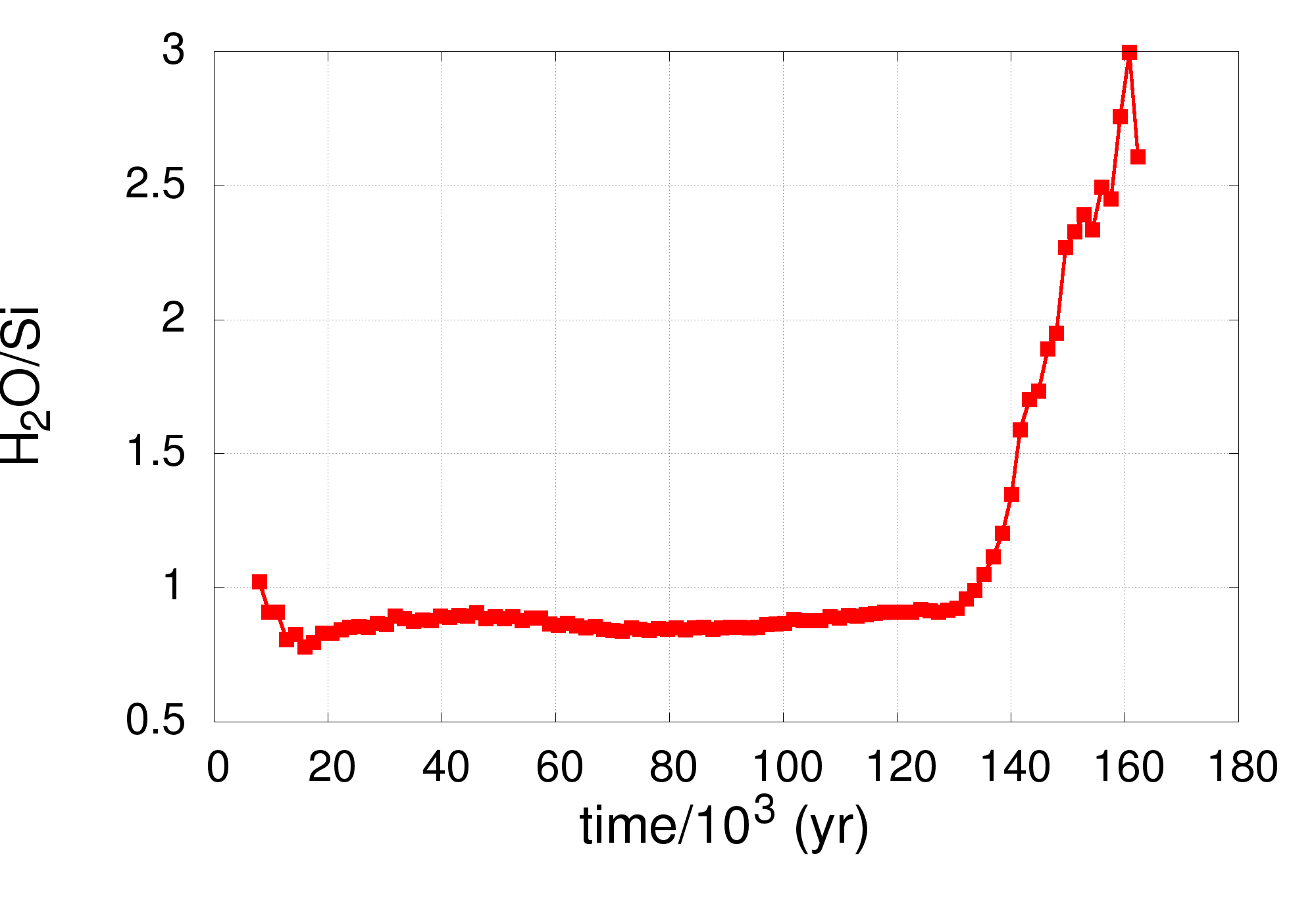}} 
{\includegraphics[width=0.65\columnwidth]{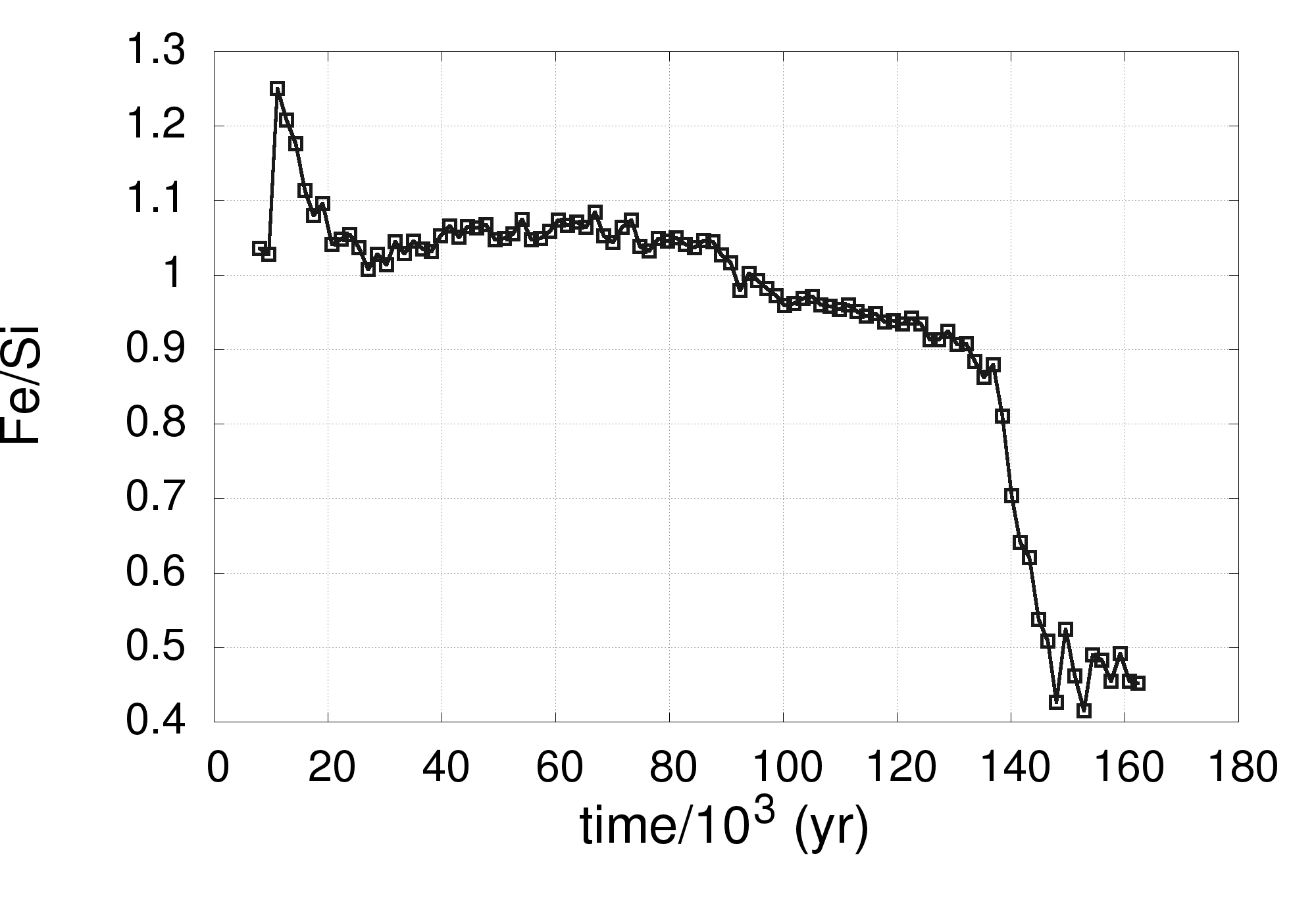}} \\
{\includegraphics[width=0.65\columnwidth]{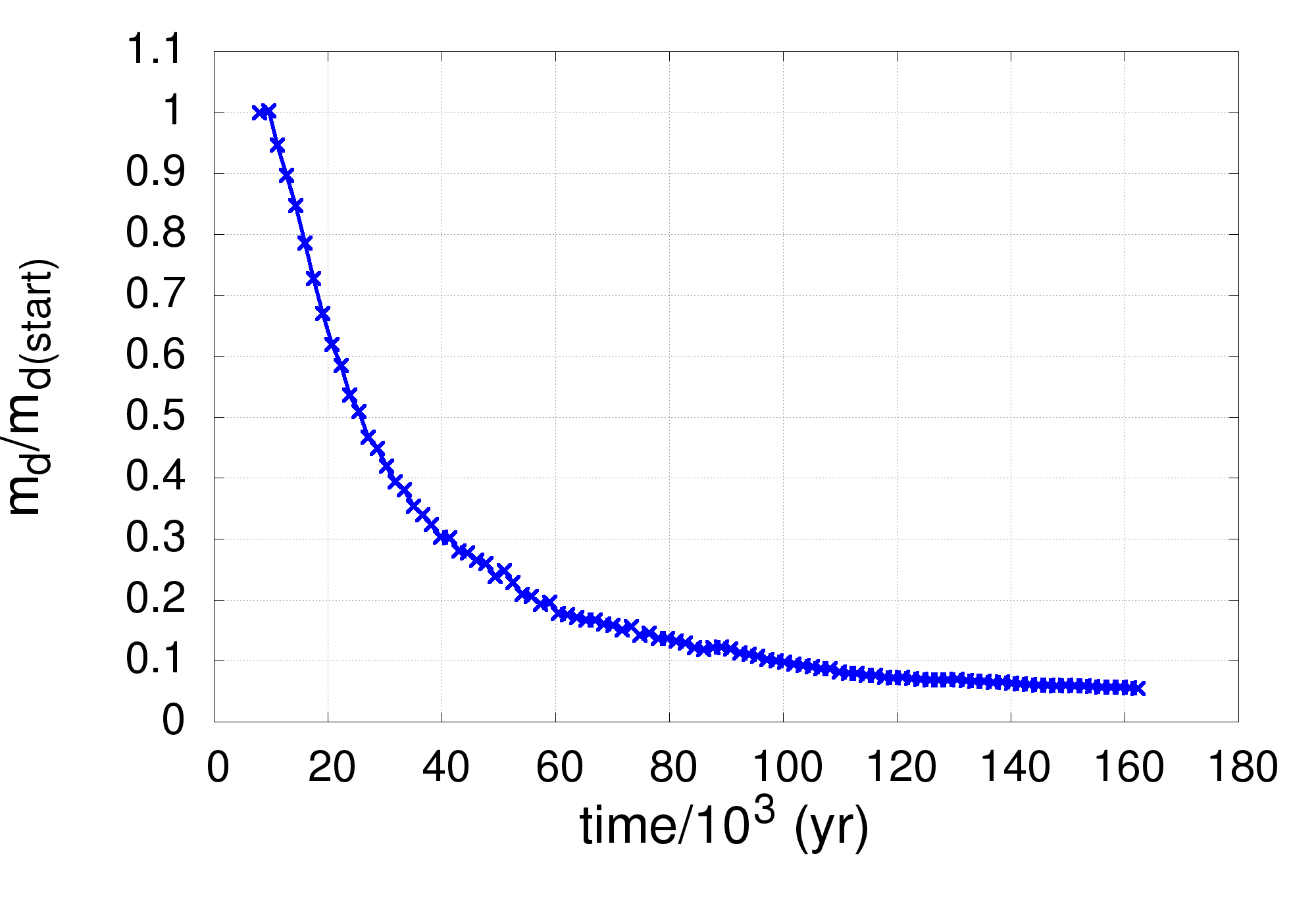}}
{\includegraphics[width=0.65\columnwidth]{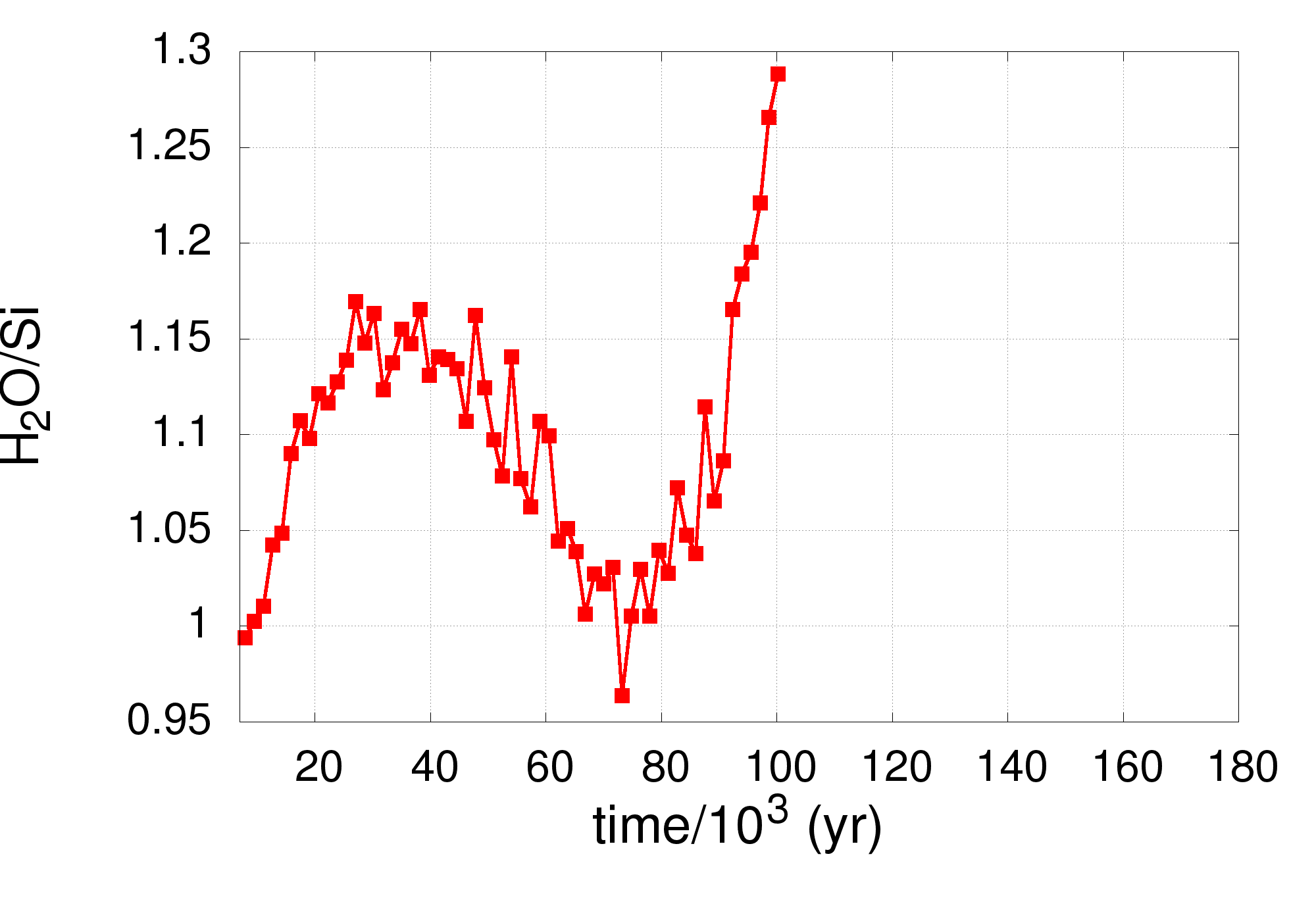}} 
{\includegraphics[width=0.65\columnwidth]{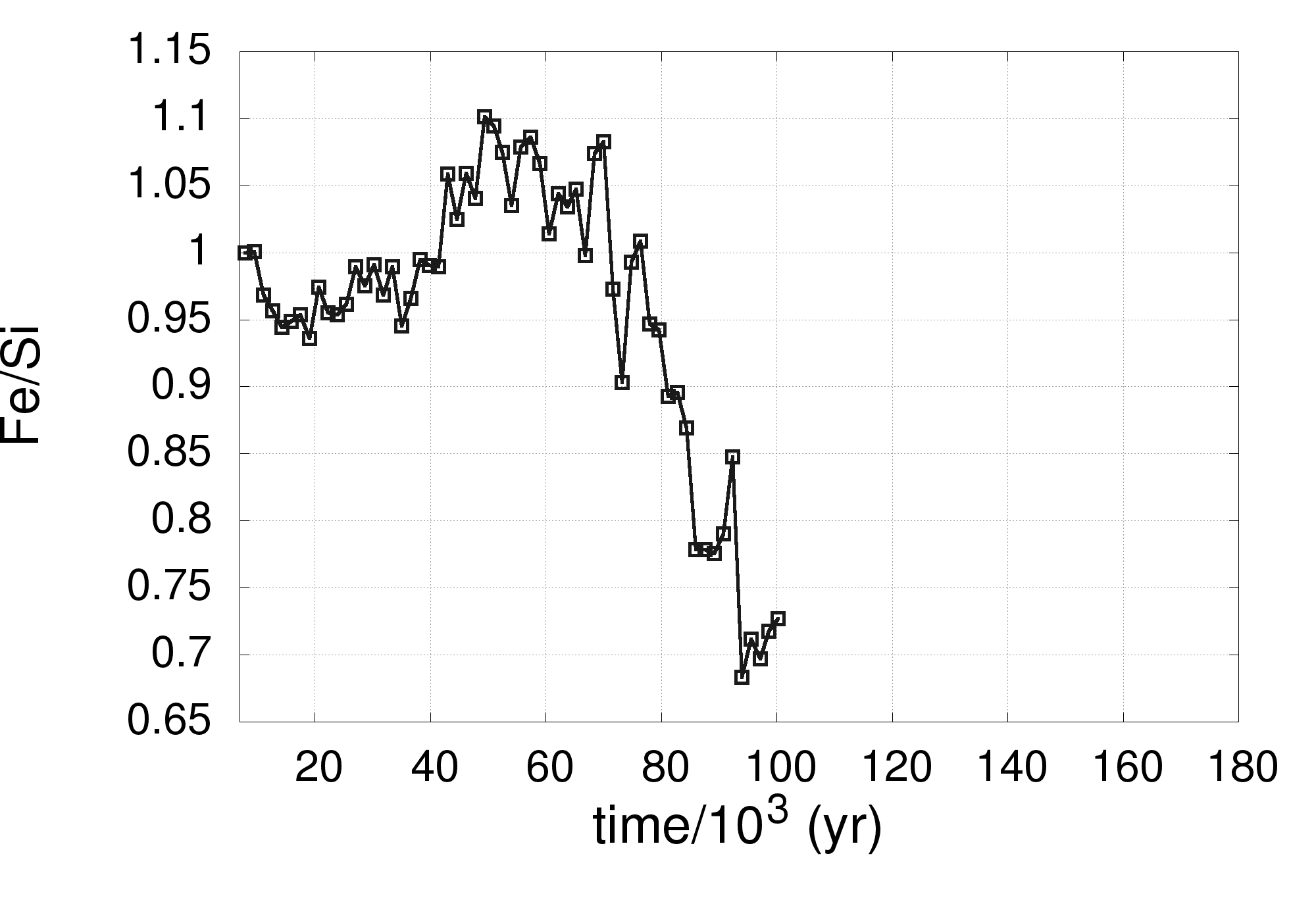}} \\
\caption{Same as Fig.\ref{100} with $100<R\rm{(AU)}<200$, in the midplane, top row, and surface, bottom row. Letters in the dust mass plot describe the different phases of vertical settling and radial drift (see text).\label{200}}
\end{figure*}

In order to quantify in more details the change in the chemistry in the whole time range, we report in Fig.~\ref{100}, from left to right, the evolution of the  dust mass content (normalized to the mass present at the beginning of the simulation), the  evolution of the \ce{H2O}/\ce{Si} and  \ce{Fe}/\ce{Si}  ratios\footnote{As pointed out in the introduction, these ratios are the ratios between the number of particles belonging to two of the different groups (Fe, Si, \ce{H2O}) which are populating a given region at a given time. For example, the  \ce{Fe}/\ce{Si} ratios is
\begin{equation}
 Fe/Si=(n_{\rm iron}/n_{\rm sil}).
\end{equation}
All values are then normalized to the ``solar" values in Table~\ref{frac-abundance}.
}
(all values normalized to our ``solar" values in Table~\ref{frac-abundance}), in  the disc midplane ($-1<Z\rm{(au)}<1$) (top row), and in the disc surface where $\lvert Z\rm{(au)}\rvert>1$  (bottom row). The selected disc radial extension is $20<R\rm{(au)}<100$.

The \ce{H2O}/\ce{Si} and  \ce{Fe}/\ce{Si} ratios at the surface of the disc are plotted until the dust mass, m$_{\rm d}$, drops to 10\% of the initial mass. Close to this value the  ratios are biased by the small number of particles present. The time resolution of all the plots is $\Delta t\sim159$~yr. This allows to investigate the  variation with time of the chemical content in our disc in very good detail, as this value represents 1/1000 of the total time of the simulation. Moreover, these plots will give us an idea of the chemical composition of the dust from which aggregates could accrete at a given time.

Looking at Fig.~\ref{100}, the effects of the dynamics on the chemical distribution of the dust become clearer.  For the  \ce{H2O}/\ce{Si} ratio, in the midplane, we see (1) the enhancement of silicate particles due to the density-driven vertical settling; (2) the size-driven vertical settling then increases the amount of ice which is populating the midplane. When the dust radial drift starts to become more efficient, (3) we see that silicate particles from the outer disc are populating the inner disc midplane, thus  decreasing the \ce{H2O}/\ce{Si} ratio. At later stages, (4) the radial drift of the  ice  increases  the \ce{H2O}/\ce{Si} ratio. On the other hand, for the \ce{Fe}/\ce{Si} ratio, we see (1) the enhancement of iron due to the density-driven vertical settling; (2) the amount of silicate particles is then increased by the size-driven vertical settling. When the dust radial drift starts to become more efficient (3) iron particles populate the inner disc midplane, and, at later stages, (4) the radial drift of the  silicates decreases the \ce{Fe}/\ce{Si} ratio.

At the same time, on the surface of the disc, the dust content is reduced in heavier compounds.  The  \ce{H2O}/\ce{Si} ratio increases due to the vertical settling of the silicate particles. The trend does not change with time as the the vertical settling of the icy particles does not become so efficient to invert and lower the \ce{H2O}/\ce{Si} ratio  within the time ($t\sim20$~kyr) for which we reach the cut-off (the mass of the particles present in the selected zone reaches 10\% of the initial mass). The \ce{Fe}/\ce{Si} ratio becomes ``sub-solar" due to the vertical settling of the iron particles, until the density-driven phase is replaced by the size-driven phase when  silicate particles also start to settle more efficiently due to their growth. As a consequence  the \ce{Fe}/\ce{Si} ratio starts to move again towards higher values.

Similary to Fig.~\ref{100}, in Fig.~\ref{200} we report the dust mass ratio and the chemical ratios in the disc midplane (top) and surface (bottom) for $100<R\rm{(au)}<200$. We see a different behaviour of the mass of particles which populate the disc midplane at different times: (a) the amount of dust in the midplane increases due to the dust vertical settling, and  (b) at $t\sim50$~kyr due to the radial drift of the dust coming from the outer regions of the disc. The behaviour of the chemical ratios at these early stages is similar to that found in Fig.~\ref{100}. However, at $t\sim80$~kyr we see (c), a steep decrease of the dust mass of the disc and, at the same time, a decrease of the  \ce{Fe}/\ce{Si}  ratio. At  $t\sim120$~kyr, (d), a second fast depletion of dust occurs and the  \ce{H2O}/\ce{Si} increases.  This behaviour can be explained by the radial drift of the iron particles which starts as soon as their optimal drift size  is reached (c), and then by the radial drift of silicates particles, when they reach their $s_{\rm opt}$ at later stage (d) (see also Fig.~\ref{trackgrowth}). 

On the disc surface  the chemical ratios are characterized by the similar behaviour found in Fig.~\ref{100} for the early evolutionary stage: in this part of the disc surface, the \ce{Fe}/\ce{Si} ratio drops to ``sub-solar" values (density-driven settling) before returning to ``super-solar" values (size-driven settling). The \ce{H2O}/\ce{Si} ratio increases because of the density driven vertical settling of the silicates, then decreases because of the size driven vertical settling of the ice particles. 

The second drop  in the \ce{Fe}/\ce{Si} ratio which starts at $t\sim60-80$~kyr, is then due to the radial drift of the iron which at this stage is more efficient than the silicates one, and the \ce{H2O}/\ce{Si} ratio increases because of the radial drift of the silicates which is more efficient than the ice one. However, at this stage, the total mass in this region already dropped under $20\%$ of the initial value and the results start to get biased by the small number of particles.

In conclusion, when we look in this disc zone, we are also observing the transit of the denser dust which is leaving the outer regions of the disc and crossing this zone to populate the inner disc.

\subsubsection{Analogies with chondrites}
\label{when}

In the previous sections we showed that at early stages, when the size of the pristine dust is comparable, the density-driven vertical settling will separate the denser particles from the lighter particles leading to dust chemical sorting. Our results also show that chemical sorting can also occur via radial drift: the resulting inner disc is enriched with denser material while the outer disc is characterized by a high ice content. We also showed that the combined effects of vertical settling and radial drift  continuously change the chemical content  of the disc.

Physical sorting and grain segregation are thought to be possible mechanisms which  produced dust fractionation in pristine grains in the Solar nebula and the metal-silicate fractionation observed in meteorites \citep{1999Icar..141...96K,2006E&PSL.248..650Z}. This sorting may have been one of the mechanisms which produced the different chemical compositions among the different clans of chondrites and the differences found among chondrites groups belonging to the same clan \citep{Palme200341,2005ASPC..341..953W,2005ASPC..341...15S}. 
 
However, there is still a debate about the efficiency of aerodynamic sorting and on the location at which  dust chemical sorting might have occurred. It has been suggested that different evolutionary times and radial distances from the forming Sun had played a major role in determining the final structure of the chondrites parent bodies \citep{2005ASPC..341...15S}.

In our simulation we are considering a large disc, thus we cannot directly compare our results with the chemical and dynamical evolution  of the chondrites which formed in the inner Solar Nebula. However, an indirect comparison within the frame of our disc may help to find some analogies and determine if, when and where aggregates in our disc model can resemble the major properties of chondrites, and thus, suggest pathways of formation for these objects. 

In the following discussion we consider silicate and iron particles in our disc respectively as the possible chondrules precursors (or already formed chondrules) and the metallic grains in chondrites. The mechanism (or mechanisms) from which chondrules originated is still not well constrained (shock waves, planetesimals impacts, lightning, jets, winds \citep{2005ASPC..341...15S}), and more recent analysis suggests that chondrules formation may have started  contemporaneously with CAIs (the oldest objects in our solar system) in a continuous process which lasted $\sim$3 Myr \citep{2012Sci...338..651C}. As such it is not unreasonable to consider the presence of a sparse population of chondrules in the disc since the early stages of the protoplanetary disc evolution.

Our results show that radial chemical sorting characterizes the disc only when  large scales are taken into consideration. The \ce{Fe}/\ce{Si} ratio in our inner midplane $20<R\rm{(au)}<100$ is always  barely ``super-solar" with a deviation close to 1\% (see Fig.~\ref{100}), while the  \ce{Fe}/\ce{Si}  of chondrites moves from solar (ordinary chondrites) toward sub-solar values \citep{2006mess.book..803R}  (see  Fig.~\ref{fractionation}). Thus, aggregation of ``sub-solar" material in this location can be excluded.

 We found that the \ce{Fe}/\ce{Si} in the  midplane where $100<R\rm{(AU)}<200$ (Fig~\ref{200}) is ``super-solar" until later stages (t$\sim$80~kyr) when the fast radial drift starts to deplete the denser dust in this region.   In this case assembling  dust in this part of the midplane might lead to chemically fractionated aggregates, with ``sub-solar" \ce{Fe}/\ce{Si} ratios.

However, in order to produce fractionation in the outer midplane,  time scales in the order of 60~kyr are needed to lower the \ce{Fe}/\ce{Si} ratio from 1 to 0.9 (see  Fig.~\ref{200}). This is in case of a clean and efficient radial drift. Indeed,  the radial drift of dust, and thus, the chemical sorting, can be slowed down and/or stopped by the presence of particle traps in the disc. Particles traps are thought to be  locations at which a fast and efficient pile-up and growth of dust occurs, making the formation of larger planetesimals more favourable. 

Particle traps can occur from the early stages of the protoplanetary disc evolution within vortices \citep{1995A&A...295L...1B,2004A&A...417..361J}, at the ice-lines and at the edge of the dead zone \citep{2007ApJ...664L..55K, 2008A&A...487L...1B}, in  discontinuities of the gas distribution in discs \citep{2012A&A...538A.114P} or in self-induced dust traps \citep{2015MNRAS.454L..36G,2017MNRAS.467.1984G}. Particle traps can, thus, slow down or totally stop the radial chemical separation of the dust.

Vertical settling produces instead dust fractionation all along the disc surface (see Figs.~\ref{100},and~\ref{200}). The disc surface is the only region in which we find ``sub-solar" values of \ce{Fe}/\ce{Si} ratio from the beginning of our simulation. Aggregates in the surface of the disc can have ``solar" values of the \ce{Fe}/\ce{Si} ratio at the beginning of the simulation, and ``sub-solar" values a few thousands years later.

Furthermore, grain growth experienced by particles, and described in  section~\ref{graingrowth}, suggests that aggregates with different \ce{Fe}/\ce{Si} ratios could also contain grains with different sizes.
\begin{figure*}
{\includegraphics[width=1\columnwidth]{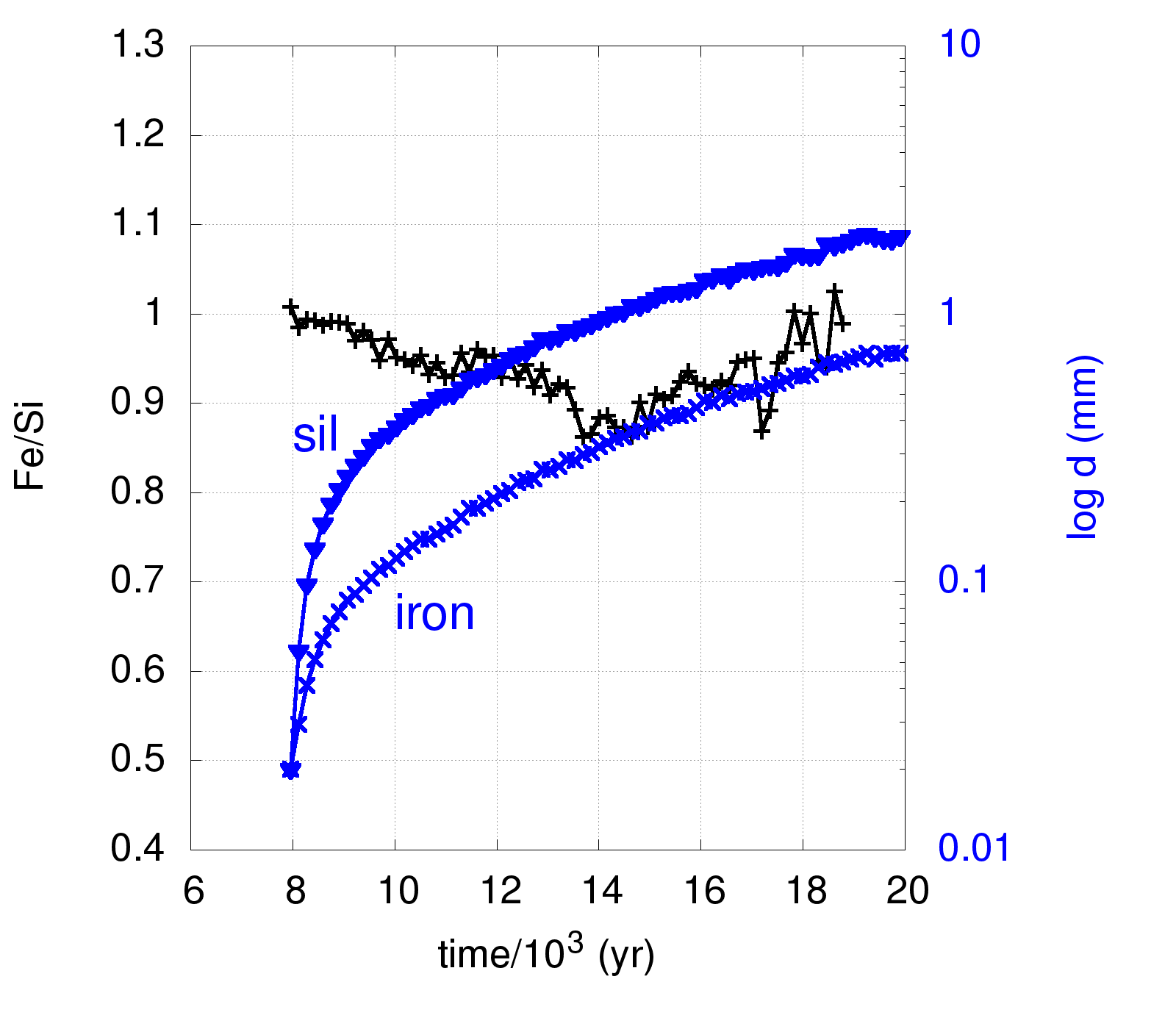}}
{\includegraphics[width=1\columnwidth]{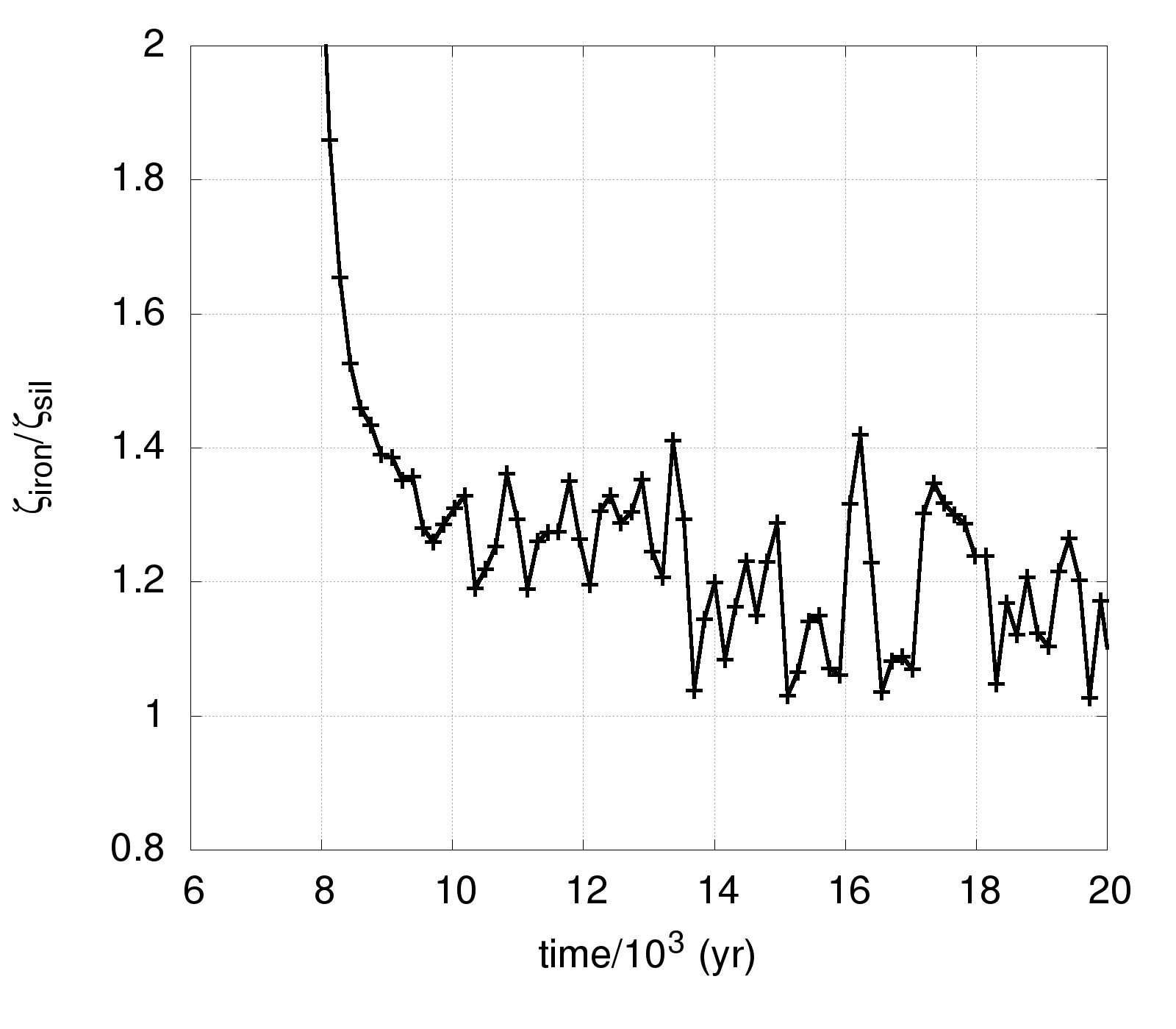}} \\
{\includegraphics[width=1\columnwidth]{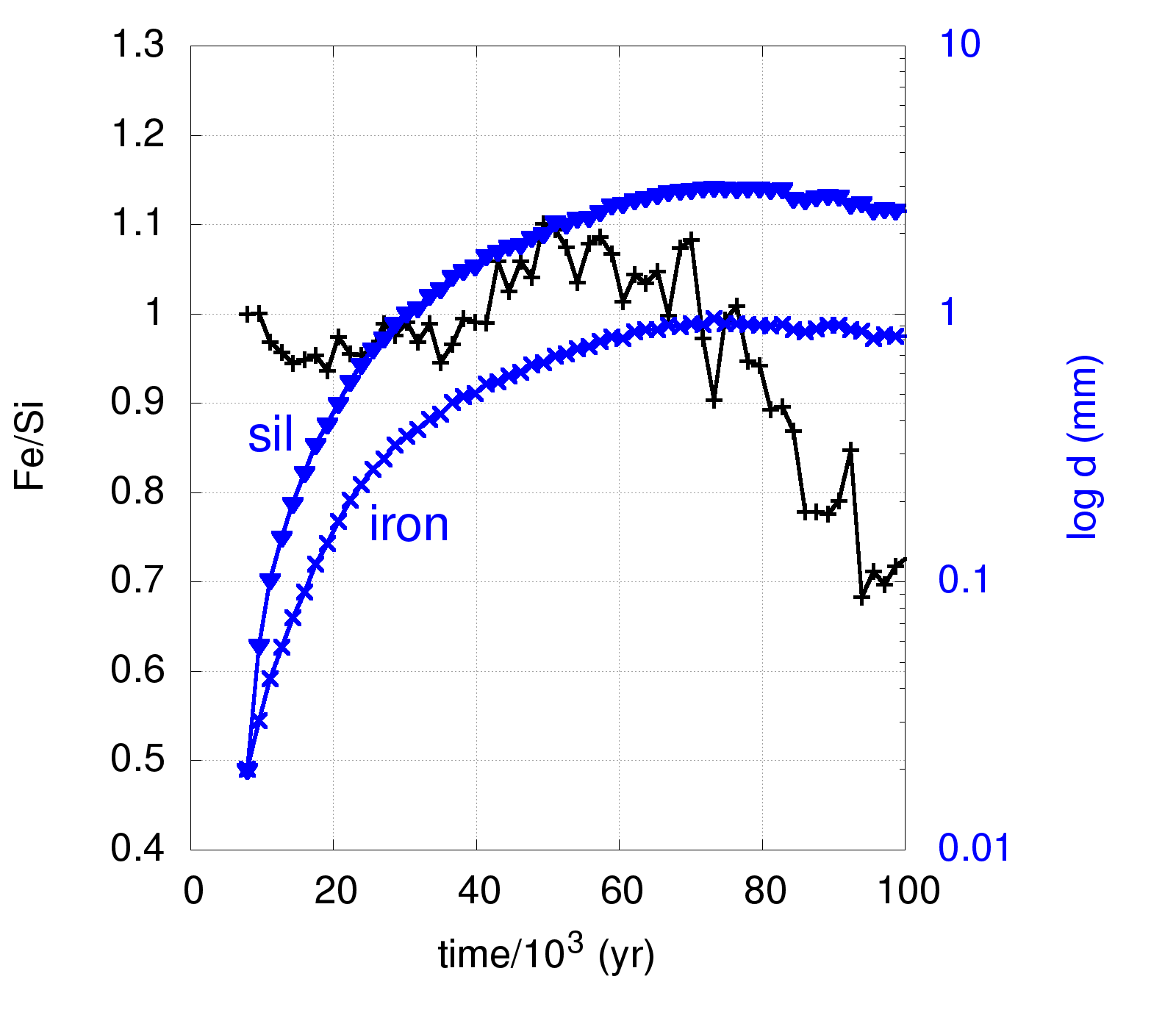}} 
{\includegraphics[width=1\columnwidth]{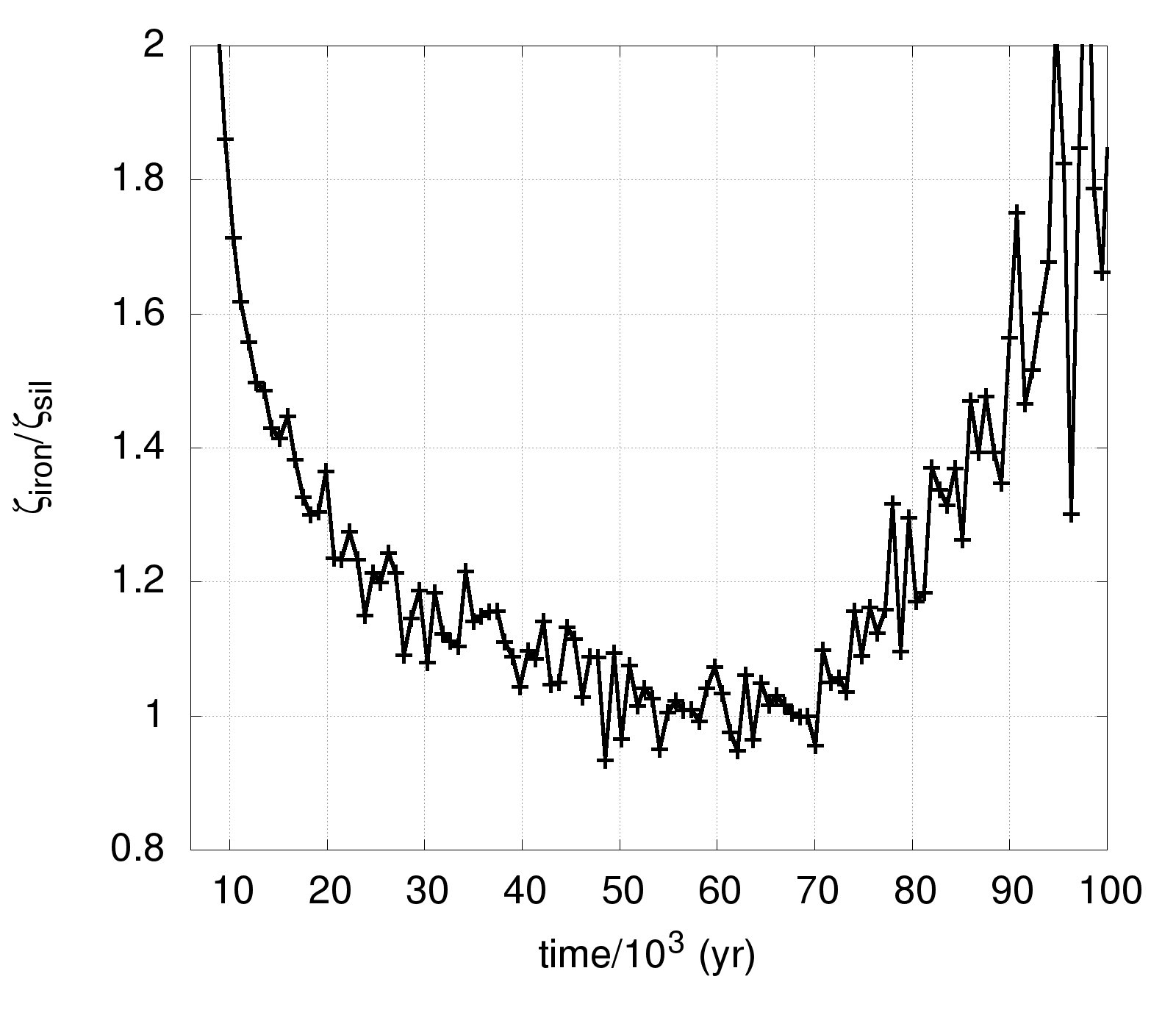}} \\
{\includegraphics[width=1\columnwidth]{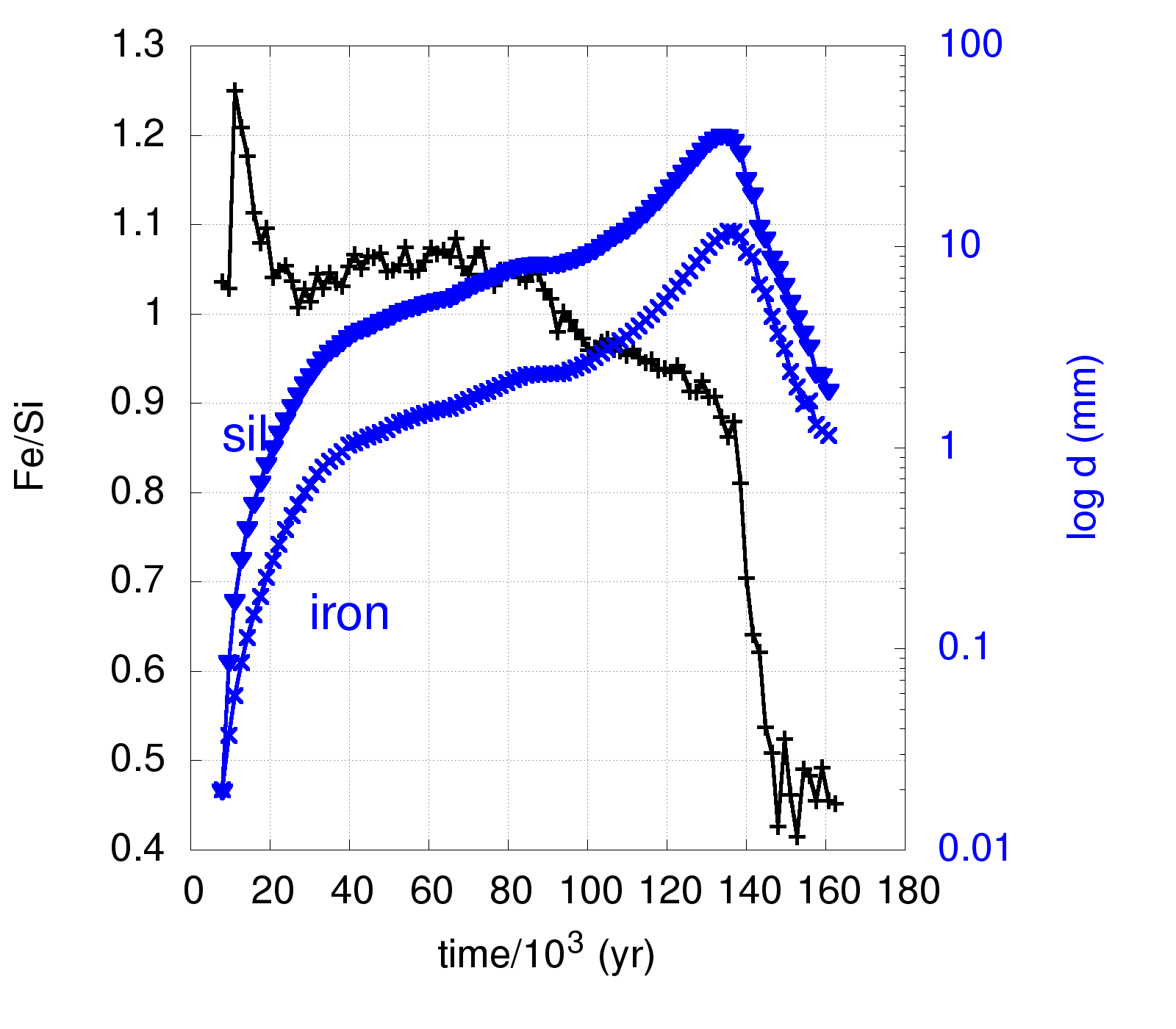}} 
{\includegraphics[width=1\columnwidth]{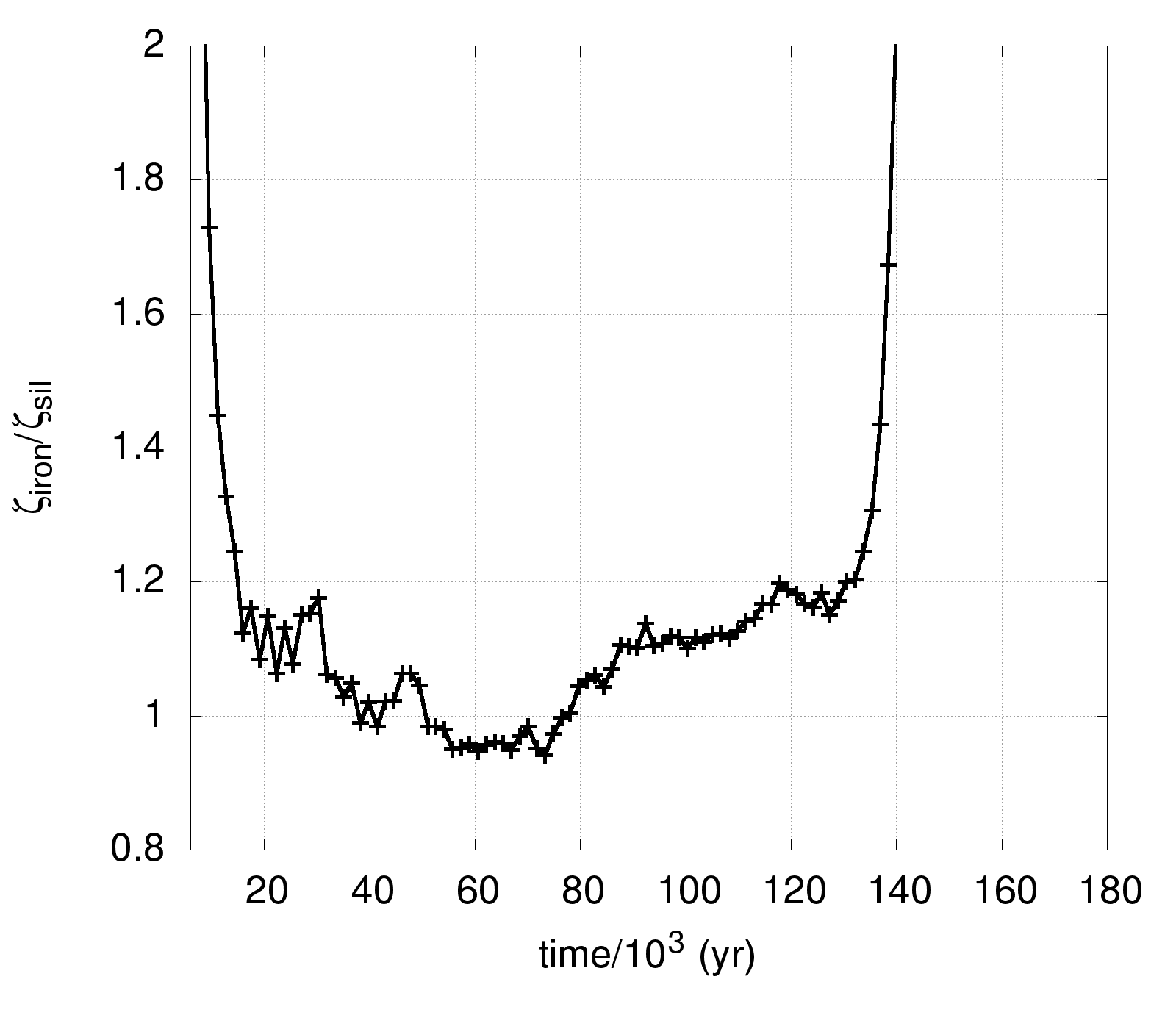}} \\
\caption{Left column:  \ce{Fe}/\ce{Si} ratio (black) and average diameter of iron and silicate particles (blue) as a function of time. Top: disc surface ($\lvert Z\rm{(au)}\rvert>1$) where $20<R\rm{(au)}<100$. Middle: disc surface where $100<R\rm{(au)}<200$.  Bottom: disc midplane ($-1<Z\rm{(au)}<1$) where $100<R\rm{(au)}<200$. Right column: $\zeta_{\rm iron}/\zeta_{\rm sil}$ for the three selected zones, calculated multipling the average size of iron and silicates with their respective intrinsic densities. A perfect size-density sorting would occur when these ratios are equal to 1. The $\zeta_{\rm iron}/\zeta_{\rm sil}$ ratios decreases quickly towards 1 suggesting that the system evolves towards size-density sorting. However, in the outer surface, after 70~Kry, and in the midplane, after 80~kyr, when the  \ce{Fe}/\ce{Si} ratio moves to ``sub-solar" values, the $\zeta_{iron}/\zeta_{sil}$ moves away from a perfect size-density sorting. \label{sizechem}}
\end{figure*}
 In Fig.~\ref{sizechem} we superimpose, on the left column, the \ce{Fe}/\ce{Si} ratio and the average diameter of silicate and iron particles as a function of time, for the disc surface ($\lvert Z\rm{(au)}\rvert>1$) where $20<R\rm{(au)}<100$ (top), where $100<R\rm{(au)}<200$ (middle), and for the disc midplane ($-1<Z\rm{(au)}<1$) where $100<R\rm{(au)}<200$ (bottom). In the right column we trace the ratio between the aerodynamic parameters of iron and silicate particles, $\zeta_{\rm iron}/\zeta_{\rm sil}$, as a function of time. The aerodynamic parameters are calculated using the average size of silicate and iron particles multiplied by their respective intrinsic densities. The closer the value of $\zeta_{\rm iron}/\zeta_{\rm sil}$ is to 1, the closer to a perfect size-density sorting the particles are. Thus, Fig.~\ref{sizechem} allows to simultaneously check the chemical fractionation of the dust, the average grains diameter, and the  aerodynamical sorting of grains in three different zones of the disc and at each time step of our simulation.

We see that aggregates with lower \ce{Fe}/\ce{Si} ratio will also have bigger grains.  Furthermore, Fig.~\ref{sizechem} (right column) shows that as soon as the simulation starts, the $\zeta_{\rm iron}/\zeta_{\rm sil}$ ratios move quickly towards values indicating a good degree of size-density sorting.

In the disc surface where $20<R\rm{(au)}<100$,  values of $\ce{Fe}/\ce{Si}<1$ are associated with grains diameter in the order of 0.1$\sim$1 mm and a $\zeta_{\rm iron}/\zeta_{\rm sil}$ value which can reach $\sim$1.2--1.1 over timescales ranging between 8 and 18~kyr. In the disc surface where $100<R\rm{(au)}<200$ we see a  $\ce{Fe}/\ce{Si}<1$ ratio for the first 30~kyr,  with grains having an average diameter in the order of 0.1 to 1~mm and with $\zeta_{\rm iron}/\zeta_{\rm sil}$ which can reach $\sim$1.1. At $t\sim$70~kyr, the  \ce{Fe}/\ce{Si} ratio drops a second time under 1, the grain diameter is in the order of mm and the $\zeta_{\rm iron}/\zeta_{\rm sil}$ ratio moves toward higher values. At this evolutionary stage, the total dust mass already dropped under $20\%$ of the initial amount (see Fig.\ref{200}), and thus, we are starting to sample the residual dust on the disc surface which did not grow/sort efficiently.

In the midplane where  $100<R\rm{(au)}<200$, after $t\sim$80~Kyr, when the  \ce{Fe}/\ce{Si} ratio moves to subsolar values, the $\zeta_{\rm iron}/\zeta_{\rm sil}$ moves away from a perfect size-density sorting. However, the values of the $\zeta_{\rm iron}/\zeta_{\rm sil}$ ratios, when $80<t\rm{(kyr)}<140$,  are still compatible with a good degree of  aerodynamical sorting ($1\le\zeta_{\rm iron}/\zeta_{\rm sil}\le1.2$). Here, the average diameter of the iron and silicate particles when the \ce{Fe}/\ce{Si} starts to become ``sub-solar" is in the order of mm to cm (see Fig.~\ref{sizechem}~bottom). Aggregates in this region would have bigger grains (mm to cm), experience chemical fractionation and a general aerodynamical sorting. Then, as already shown in Fig.\ref{icesilicatesorted}, at the end of the simulation, the residual  aggregates in the outer discs will be  less aerodynamically sorted although fractionation in their  \ce{Fe}/\ce{Si}  persists.

\begin{figure*}
{\includegraphics[width=1\columnwidth]{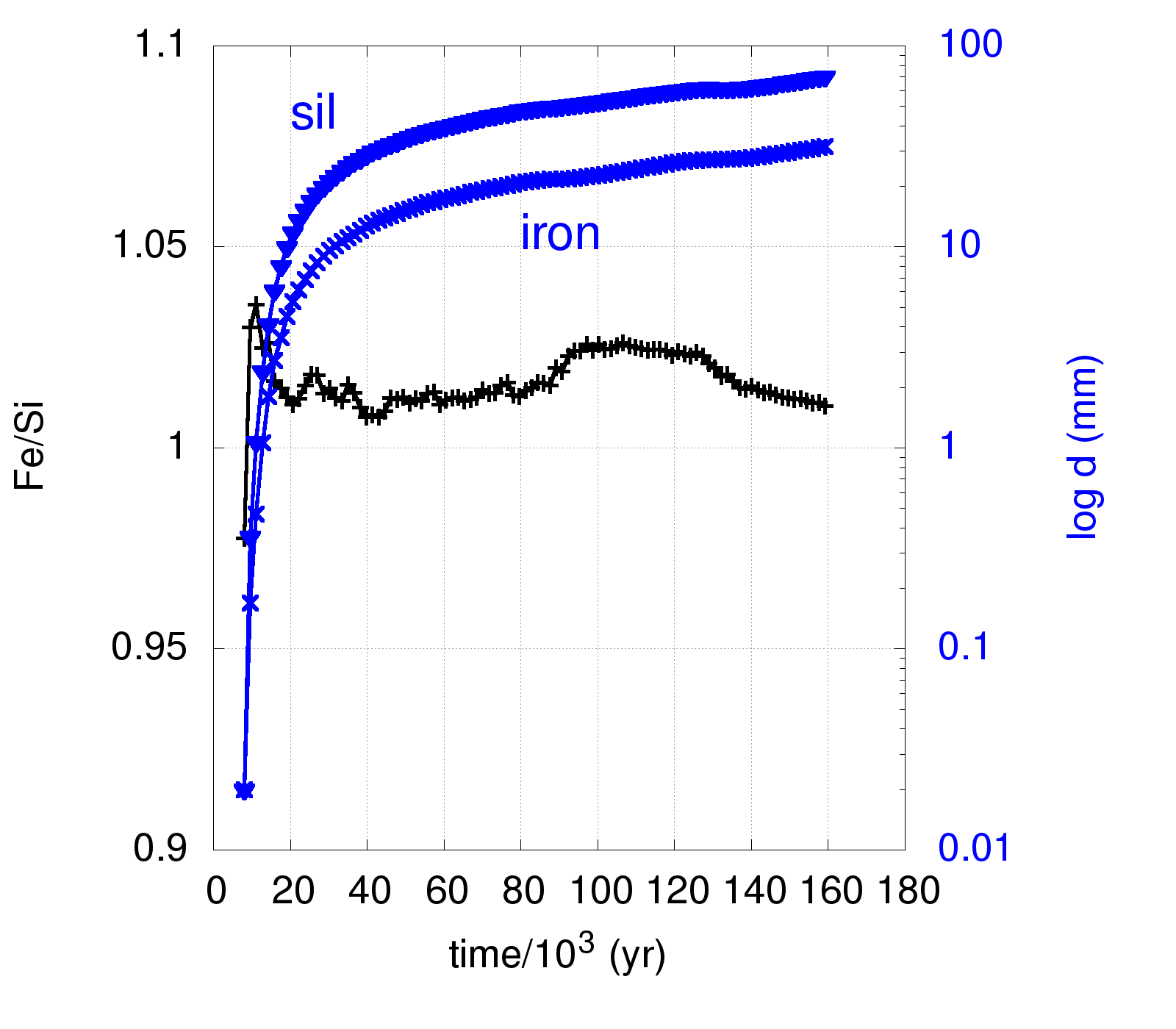}}
{\includegraphics[width=1\columnwidth]{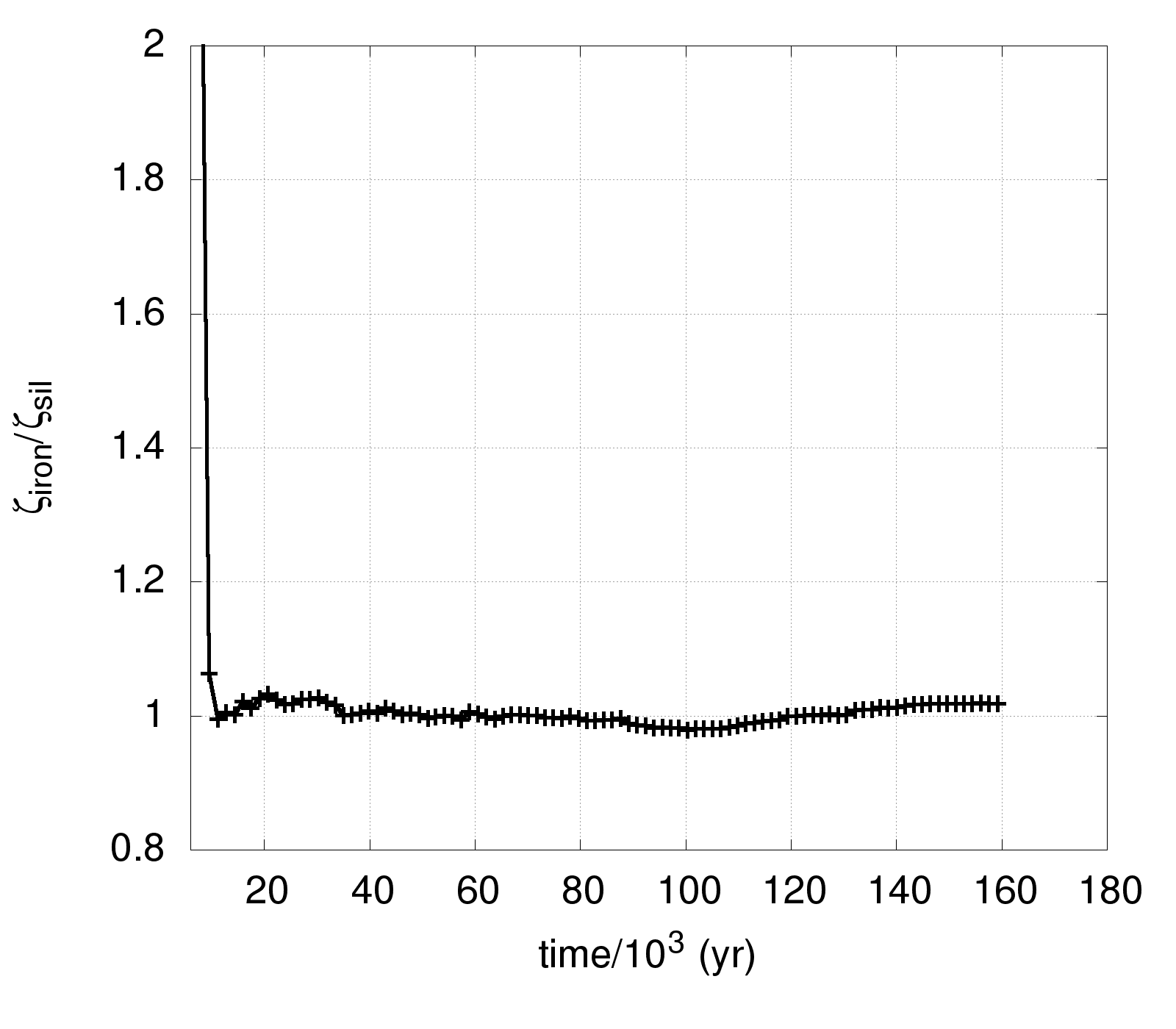}} \\
\caption{Left column:  \ce{Fe}/\ce{Si} ratio (black) and average diameter of iron and silicate particles (blue) as a function of time for the disc midplane ($-1<Z\rm{(au)}<1$) where $20<R\rm{(au)}<100$. Right column: $\zeta_{iron}/\zeta_{sil}$ for the selected zone, calculated multipling the average size of iron and silicates with their respective intrinsic densities. A perfect size-density sorting would occur when these ratios are equal to 1. The $\zeta_{iron}/\zeta_{sil}$ ratio move quickly towards 1 suggesting that the in this zone of the disc, grains move tovard a perfect sorting within few years after the beginning of the simulation. \label{sizechemmid}}
\end{figure*}

For completeness, we report in Fig.\ref{sizechemmid} the  \ce{Fe}/\ce{Si} ratio, the average diameter of iron and silicate particles and the $\zeta_{\rm iron}/\zeta_{\rm sil}$ ratio for the inner midplane where  $20<R\rm{(au)}<100$. It can be seen that particles in this zone of the disc move toward a perfect size-density sorting. The sorting in this zone is achieved almost immediately, just after the beginning of the simulation.

We thus find a  trend which is compatible with a fractionation of the \ce{Fe}/\ce{Si} ratio  and the size of the  chondrules and iron grains found in different groups of  chondrites.  However, the values of the  \ce{Fe}/\ce{Si} ratios returned by our simulations are not low enough when compared with the observed  \ce{Fe}/\ce{Si} ratios in chondrites \citep{2006mess.book..803R}. This is because of our adopted spatial resolution. When calculating our ratios in the disc surface, we are considering, for a given interval in $R$, all the vertical extension of the disc for which ($\lvert Z\rm{(au)}\rvert>1$), thus losing resolution in $Z$. 

\begin{figure*}
{\includegraphics[width=1\columnwidth]{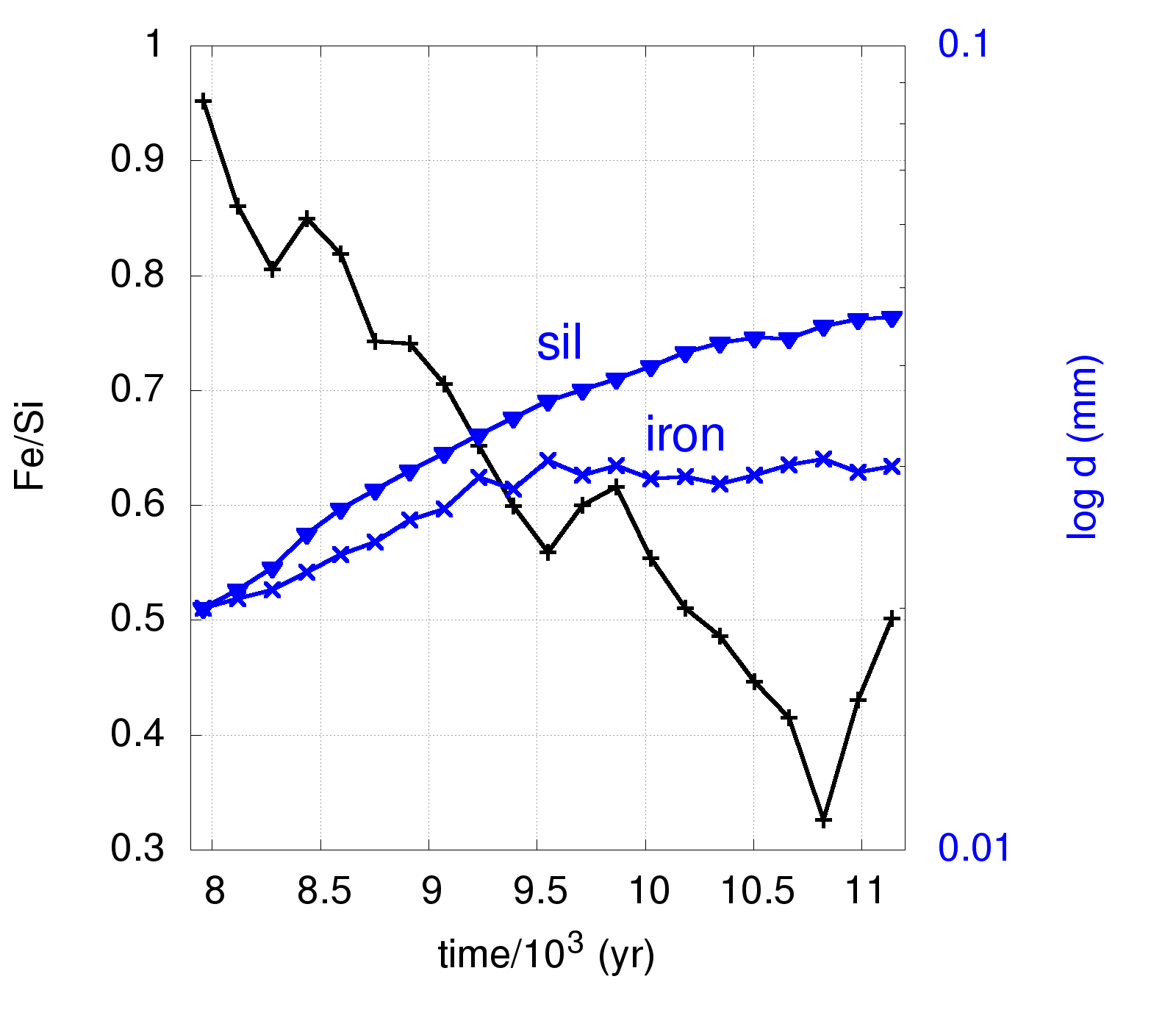}}
{\includegraphics[width=1\columnwidth]{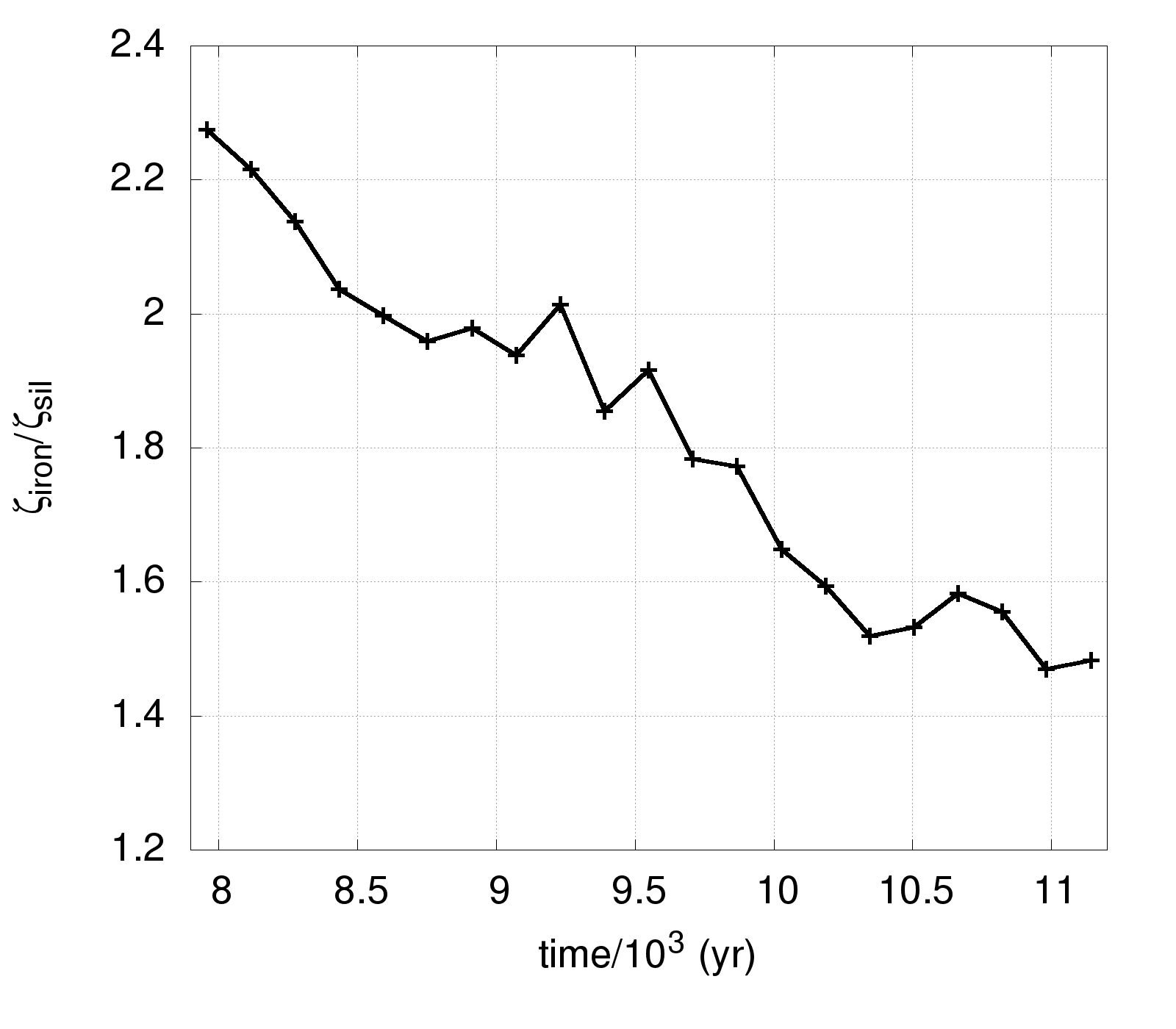}} \\
\caption{Left column:  \ce{Fe}/\ce{Si} ratio (black) and average diameter of iron and silicate particles (blue) as a function of time for the disc surface ($\lvert Z\rm{(au)}\rvert>5$) where $20<R\rm{(au)}<100$. Right column: $\zeta_{iron}/\zeta_{sil}$ for the  selected zone. \label{verysurface}}
\end{figure*}

Figure~\ref{dustevo} shows that there are zones in the disc surface for which the \ce{Fe}/\ce{Si} ratios can reach very low values, and 0 where iron particles are not present. Thus, as an example, we report in  Fig.~\ref{verysurface} the \ce{Fe}/\ce{Si} and the $\zeta_{iron}/\zeta_{sil}$  ratios for the disc surface where $\lvert Z\rm{(au)}\rvert>5$ and $20<R\rm{(au)}<100$. Ratios are plotted until the number of iron particles reaches 0.
It can be seen that, when a different vertical section of the disc is considered, the \ce{Fe}/\ce{Si} ratios can reach very low values,  more compatible with the observed trend in chondrites, although, in this case,  the size-density sorting is less efficient with a $\zeta_{iron}/\zeta_{sil}$ moving toward a minimum value of $\sim$1.4.

Furthermore, as already pointed out,  chondrites did not form at this very large disc scales, but in the inner few au of our Solar nebula. This suggest that a similar process  might have also occurred in the inner disc, with  different time-scales or with  different efficiency.

A simple extrapolation of  our results would imply that the aggregation of chondritic material could have occurred  in the surface of the inner young Solar Nebula during the two phases of vertical settling where size-density sorting was also efficient.  The differences found among chondrites clans might then be ascribed to the different radial distance at which this process occurred. This distance could have then determined, for example, the amount of water accreted from the chondrite parent bodies. Indeed accretion of E chondrites likely occurred in the inner warmer region of the disc within the snow-line, ordinary chondrites close to the snow-line and carbonaceous chondrites  beyond the snow-line \citep{2005ASPC..341...15S,2005ASPC..341..953W}. 

These results only represent a starting point in the study of the dynamics of chondrites. Indeed, more complex models are needed to explain not only the variations in the  \ce{Fe}/\ce{Si} ratios but also among other elements.

In our next study we will explore the growth and the aerodynamic sorting of a multicomponent dust in the inner disc accounting also for dust fragmentation.

\subsection{Analogies with discs observations}
\label{discanalog}

In the previous sections we showed that grains decouple from the gas and settle toward the midplane according to their density and size. Moreover, as grains reach their optimal drift size, they start to drift toward the inner region of the disc.

\begin{figure}
{\includegraphics[width=1\columnwidth]{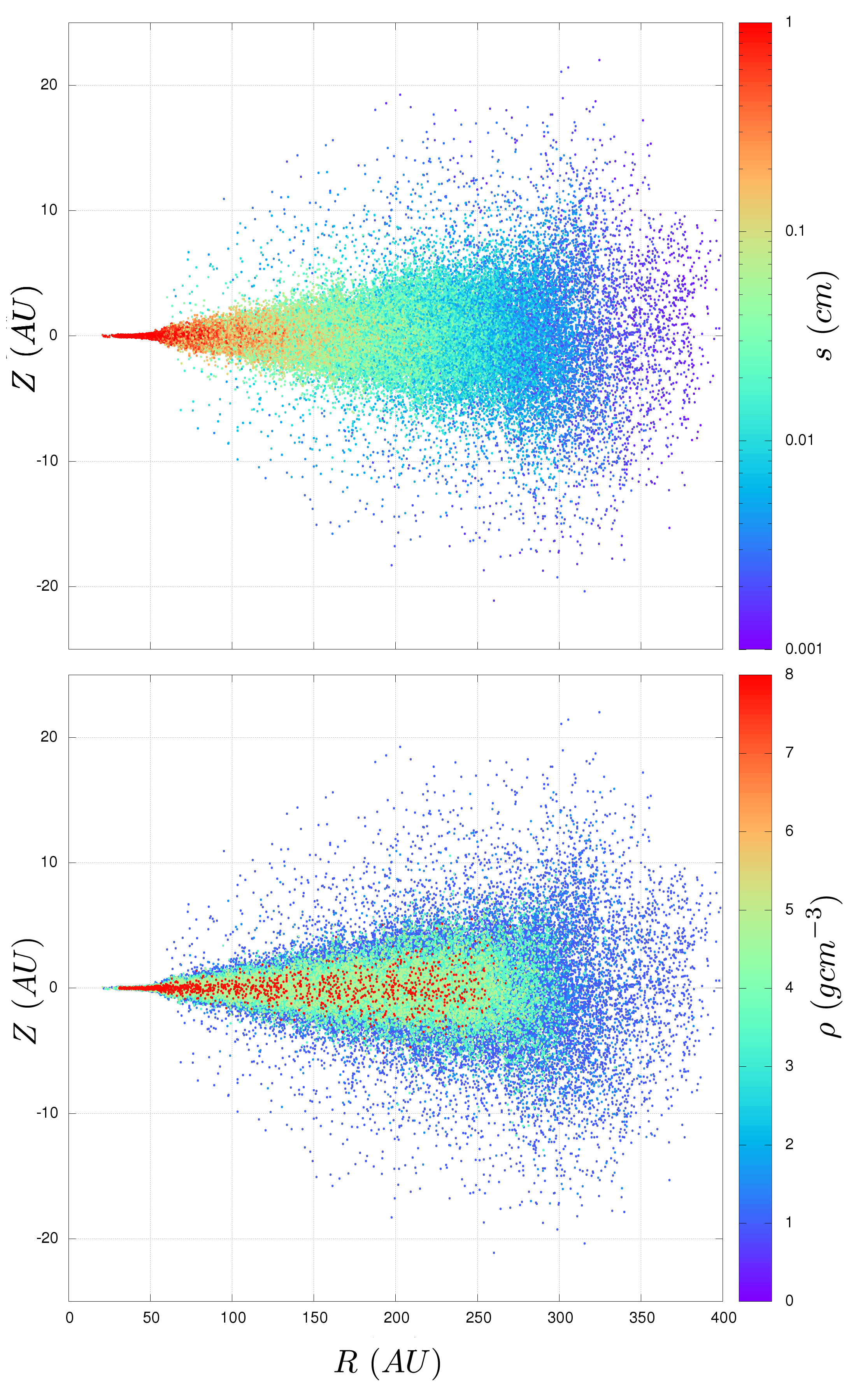}} \\
\caption{Top: Grain size distribution as a function of R and Z at $t=42970$~yr. Bottom: Distribution of different chemical species as a function of R and Z. This figure reproduces the top-right panel in Fig.~\ref{dustevo}, for ease of comparison. Larger/heavier grains are located in the inner disc midplane while smaller/lighter grains are distributed in a wider zone. This in general agreement with observations. \label{sizecolor}}
\end{figure}

In Fig.~\ref{sizecolor}(top) we plot the grain size distribution at $t=42970$~yr, as a function of $R$ and $Z$. This timestep is the same as the top-right panel in Fig.~\ref{dustevo} where we plotted $\rho_{d}$ as a function of $R$ and $Z$. That figure is reproduced in Fig.~\ref{sizecolor}(bottom) for ease of comparison.

 In terms of size, our results are in good agreement with previous simulations made by \citet{2005A&A...443..185B} who studied the dust distribution of grains with different size and found that larger grains pile up in the disc midplane while smaller grains distribute in the disc envelope. We are also in agreement with the sorting trend observed in GG Tau and TW Hya described in section~\ref{intro}.
 
Looking at the evolution of the chemical distribution in Fig.~\ref{dustevo} and comparing it with the size distribution in Fig.~\ref{sizecolor} a qualitative observable prediction of the chemistry and physical properties of the dusty grains  would read as smaller, lighter grains at the discs surface. This is in agreement with the dust properties of protoplanetary  derived by infrared observation \citep{2008ApJ...683..479B,2009A&A...497..379M} where micron-size silicates particles are used to fit the infra-red spectra of the surface of the dusty discs around young stellar objects. 

Moreover our results can help to disentangle the chemical dust composition of the disc within the midplane, which cannot be probed by observations \citep{2011ARA&A..49...67W}. We suggest that the inner disc midplane could be enriched in denser material, iron and silicate-rich aggregates, while the outer disc regions would host ice-rich and iron-poor materials (see Fig.\ref{surfdust}). This prediction is also consistent with the chemical radial gradient generally found in the solar system when moving toward larger distances from the Sun \citep{Lewis2004}: denser rocky planets in the inner regions while volatile-ice enriched bodies in the outer solar system.

Furthermore, our results suggest that planetary embryos might accrete  from  inhomogeneous material since the beginning of their formation with the location (radius) of their accretion  determining  their initial bulk elemental ratios.

\section{Conclusions}
\label{conclusion}

We presented SPH simulations of grain growth and sorting of multicomponent dust in a protoplanetary disc using a 3D, two-fluid (gas+dust) SPH code.

We found that the dust vertical settling is characterized by two phases: the first phase is driven by the dust density which leads to a vertical chemical sorting of the dust; the second phase is size-driven which allows the enhancement of the concentration of lighter material in the layers deep inside the disc. We also see an efficient radial chemical sorting of the dust, with denser material rapidly drifting in the inner zone of the disc. This process is driven by the different optimal drift size proper to each dust species.

Particles in the disc are aerodynamically (size and density) sorted in the inner regions, while in the outer regions the lower growth rates and the lower gas density prevent this sorting to occur efficiently.  Particles move toward size-density sorting in very short time-scales. Our results are compatible with the observed large scale dust sorting observed in protoplanetary discs.

We see that the growth and dynamics of a multicomponent dust clearly produce chemical fractionation and  aerodynamic sorting in large disc scales.  It is interesting to note that large discs ($20\le R~\rm{(au)} \le 400$) could produce, in a very short time (for example, t$\sim$5-20~kyr, in the inner disc surface), aggregates which would mimic the basic properties of chondrites. This suggest that dynamical chemical fractionation and aerodynamical sorting  might have played an important role in determining their final aspect and chemical composition.
 
Moreover, our results open a new interesting question on the existence of undifferentiated and chemically fractionated aggregates which might have formed in the outer Solar Nebula during the early stages of the formation of our Solar System. We also suggest that planetary embryos might be characterized by inhomogeneous composition since the beginning of their formation.


\section{Acknowledgments}
\label{acknowledgments}

The authors are grateful to the LABEX Lyon Institute of Origins (ANR-10-LABX-0066) of the Universit\'e de Lyon for its financial support within the program "Investissements d'Avenir" (ANR-11-IDEX-0007) of the French government operated by the National Research Agency (ANR). We also aknowledge funding from PALSE (Programme Avenir Lyon Saint-Etienne). The authors wish to thank Sarah Maddison for helpful discussions and the anonymous referees for their useful comments which made us investigate in more detail our assumptions and the timescales of dynamics and growth, which improved the manuscript. All simulation were performed at the Common Computing Facility of LABEX LIO.


\appendix
\section{Density versus growth}
\label{densvsgrowth}

In this appendix we investigate the efficiency of grain growth, vertical settling and radial drift. Our aim is to confirm that growth is not efficient enough to erase the inhomogeneity that different grain densities bring in the disc since the beginning of our simulations.

If grains with initially comparable sizes and different densities are located at the same position, in order to counterbalance the dynamical differences driven by  densities, lighter grains should reach the same aerodynamic parameter, $\zeta$, as denser grains  before the vertical settling and/or the radial drift separate them.

Using our disc model in its initial configuration, we first solve  equation~\ref{stepval} with a simple numerical integration \citep[see][]{2008A&A...487..265L} for three different grains  with the same $s_{0}=10~\mu$m  and different $\rho_{\rm d}$ (ice, silicates, and iron, with $\rho_{\rm d}$ taken from table~1). We assume that grains are located and locked in the same position. The chosen location is at $R$=100~au, our reference radius, and $Z=H(R)$. We  compute the time evolution of $\zeta$ and estimate the time, $t_{\zeta}$, at which grains reach the same value of the aerodynamic parameter. The gas density at $Z=H(R)$ is calculated using
\begin{equation}
\rho_{\mathrm g}(Z)=\frac{\Sigma}{\sqrt{2\pi}H}\exp\Big(\frac{-Z}{2H}\Big) ,\
\end{equation}
 in~\citet{2012A&A...537A..61L}.
Then, we repeat the calculation in the midplane at $R=100$~au. From Fig.~\ref{parttrack} it can be seen that, once the particles reach the midplane, they have grown to sizes of the order of  $100~\mu$m.  We thus use $s_{0}=100~\mu$m as our reference size for this calculation.
\begin{figure}
{\includegraphics[width=1.0\columnwidth]{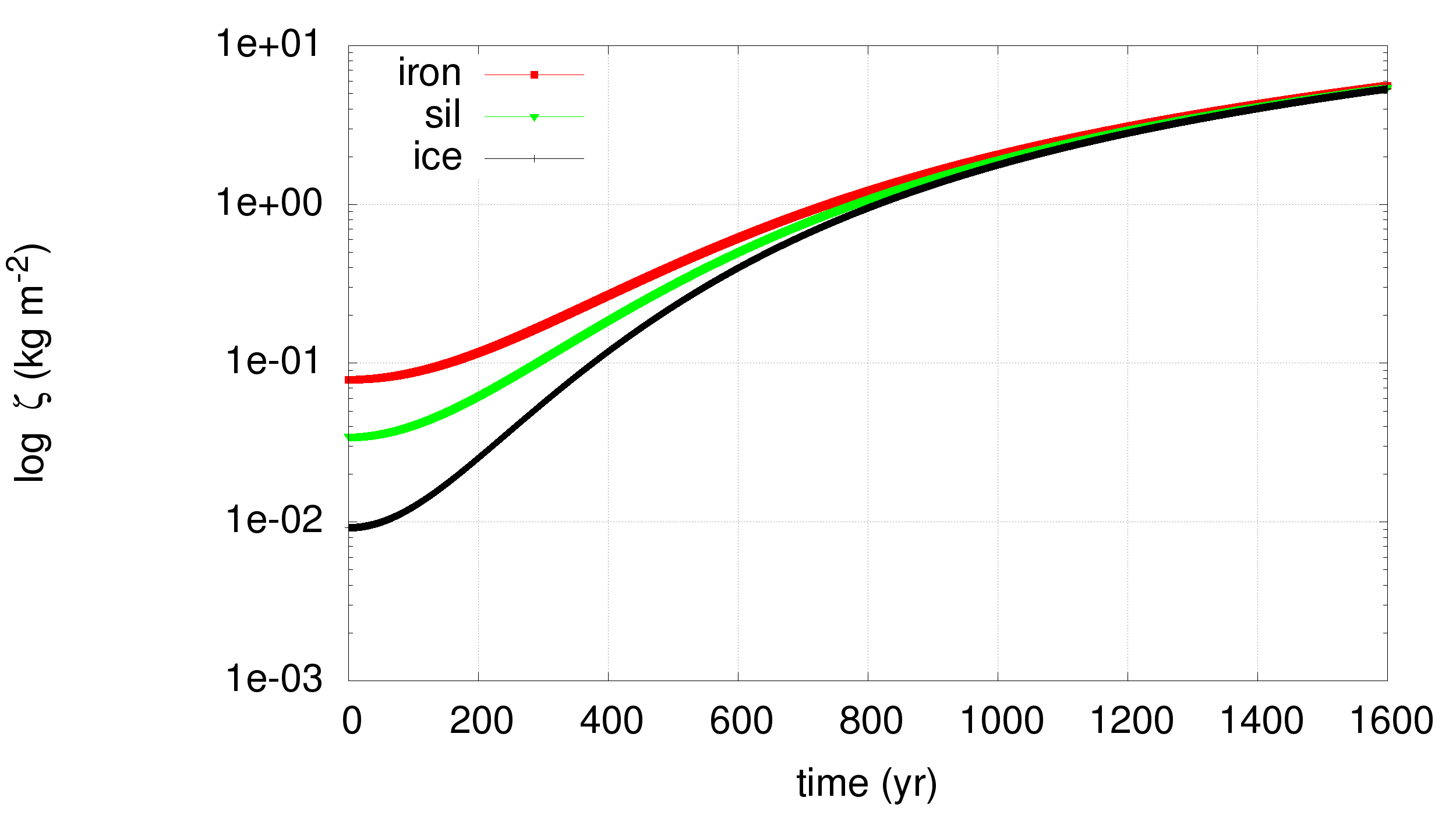}} \\
{\includegraphics[width=1.0\columnwidth]{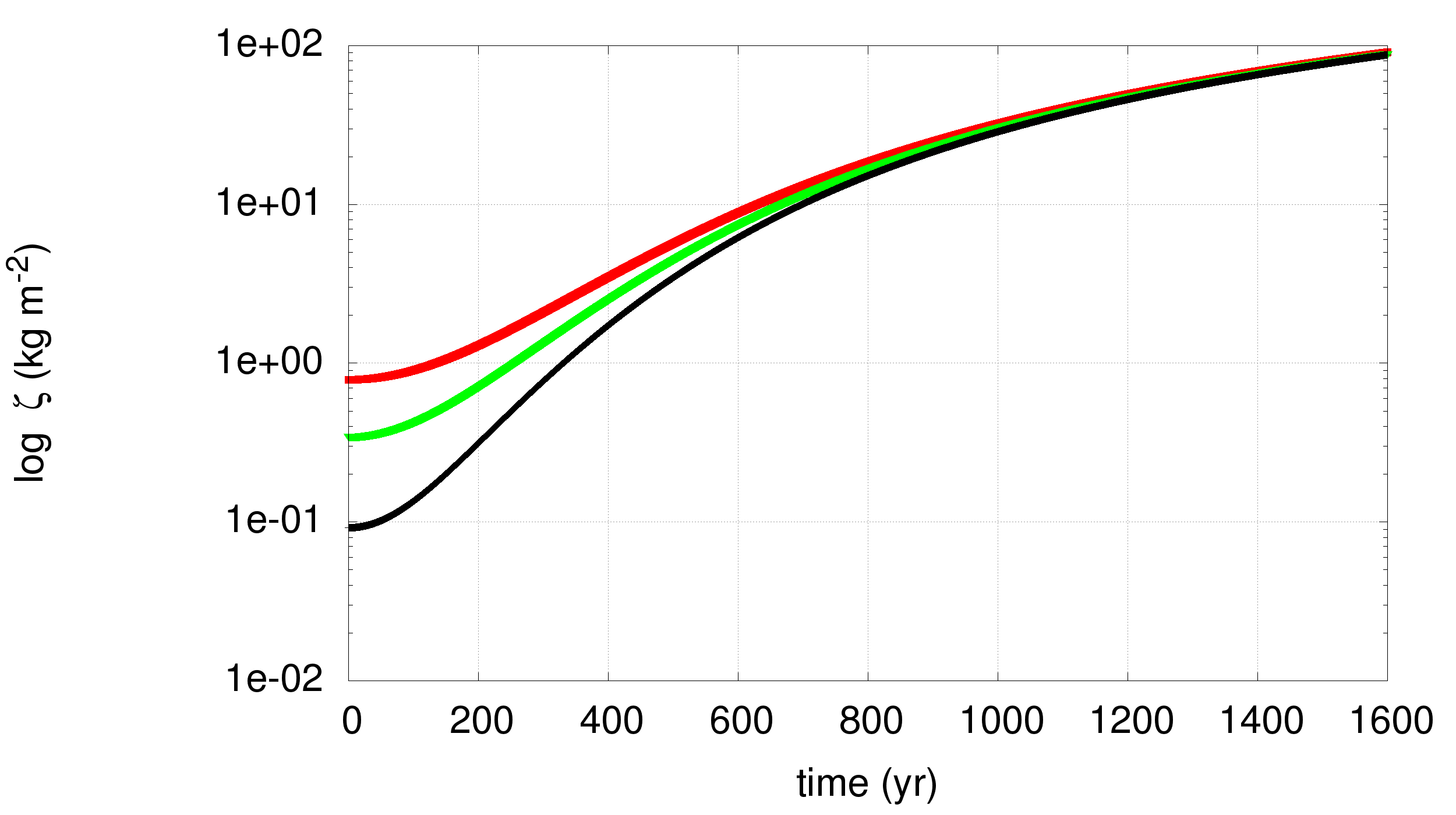}} \\
\caption{Time evolution of the aerodynamic parameter, $\zeta$, at $R=100$~au. Top, at the disc surface ($Z=H(R)$) for $s_{0}=10~\mu$m. Bottom, in the midplane for $s_{0}=100~\mu$m. \label{vermid}}
\end{figure}

Figure~\ref{vermid}  shows the  time evolution of $\zeta$  for the three different particles calculated at $R=100$~au at the disc surface ($Z=H(R)$) for $s_{0}=10~\mu$m  (top) and in the midplane for $s_{0}=100~\mu$m (bottom). Note the change of scales for $\zeta$. 
The timescales, $t_{\zeta}$, for which three particles reach a similar value of $\zeta$ is of the order of  $t_{\zeta}\sim1000$~yr.

We now calculate the initial velocity of settling, $v_{\rm set}$, and drift, $v_{\rm drift}$, proper to each grain. These quantities,  multiplied by an interval of time, $t$, give us an estimate of the separation which would occur  between grains if they were allowed to move. We set $t$=1~yr, a value much smaller than $t_{\zeta}$. If the resulting separation between grains is not negligible  we can conclude that growth cannot counterbalance the inhomogeneities that densities are bringing into the disc.

In our calculations, $v_{\rm set}$ is equal to
\begin{equation}
v_{ \rm set}=- \frac{\Omega_{\rm k}^{2}Z}{\rho_{\rm g}c_{ \rm s}}\zeta,\
\label{settequation}
\end{equation} 
derived from equation~\ref{timesettling}, which is valid for small grain sizes \citep{2004A&A...421.1075D}.

$v_{\rm drift}$ is calculated using the following equation from \citet{2008A&A...480..859B}:
 \begin{equation}
 v_{\rm drift}=v_{\rm dust}+\frac{v_{\rm gas}}{1+(\rm St^{2})} .\
 \label{driftequation}
 \end{equation}
$v_{\rm dust}$ is given by
 \begin{equation}
 v_{\rm dust}=-2\nu_{n}\frac{1}{\rm St+(1/\rm St)} ,\
 \end{equation}
where
 \begin{equation}
 \nu=\frac{c_{\rm s}^2}{2V}\Big(\frac{7}{4}+p\Big) .\
 \end{equation} 
$v_{\rm gas}$ is equal to
 \begin{equation} 
  v_{\rm gas}= -3\alpha \frac{c_{\rm s}^2}{V}\Big(\frac{3}{2}-p\Big) ,\
 \end{equation} 
where
 \begin{equation}
 V=\Omega R .\
 \end{equation}

\begin{table}
\begin{tabular}{|l|c|c|c|}
 & iron & silicate & ice \\ 
\hline 
iron & -- & \textcolor{red}{229} & \textcolor{red}{360} \\ 
\hline 
silicate & \textcolor{black}{16790} & -- & \textcolor{red}{132} \\ 
\hline 
ice & \textcolor{black}{26460} & \textcolor{black}{9670} & -- \\ 
\hline 
\end{tabular} 
\caption{Radial separation at $R=100$~au and $Z=0$, between 100~$\mu$m size iron, silicate and ice particles (red) and vertical separation at $R=100$~au and $Z$=$H(R)$, between 10~$\mu$m size iron, silicate and ice particles (black). Quantities are calculated after $t$=1~yr and distances are expressed in km. \label{driftsett}}	
\end{table}

In Table~\ref{driftsett} we report the resulting radial, $\Delta R=(v_{\rm drift}^{i}-v_{\rm drift}^{j})t$,  (red), and vertical, $\Delta Z=(v_{\rm set}^{i}-v_{\rm set}^{j})t$, (black), separation, in km, between particles with different densities, $i,j$, and the same initial size, 10~$\mu$m, for the vertical settling from the disc surface, and 100~$\mu$m for the radial drift starting in the midplane. 

Values in Table~\ref{driftsett} show that drift and settling are very efficient processes which separate the particles well before they could reach the same $\zeta$. Since all the calculated quantities scale according to the disc conditions, the separation of particles would occur at all  disc radii, but with different timescales.

These values are derived under the simple assumption that $v_{\rm drift}$ and $v_{\rm set}$ are constant in the considered small interval of time. Similarly our estimation of $t_{\zeta}$ assumes that particles grow without moving. In reality $v_{\rm drift}$, $v_{\rm set}$ and d$s/$d$t$ change with time,  local disc conditions, and  particles properties. Our code solves for all the quantities at play: in Fig.\ref{parttrack} and Fig.\ref{trackgrowth} we showed the full pathways of settling, drift and growth, of particles which are located in the same position at the beginning of the simulation, and found that they separate  quickly. 

In conclusion, we see that growth is not efficient enough to counterbalance the different dynamical behaviour brought by densities since the early evolutionary stages. Dynamics will bring chemical heterogeneities in the disc where grains in different location will accrete from different available chemical reservoirs.



\newpage
\bibliographystyle{mn2e}
\bibliography{biblio}
\end{document}